\documentclass[a4paper,11pt]{article}
\pdfoutput=1 

\usepackage{jheppub} 

\usepackage[T1]{fontenc} 
\usepackage{graphicx}
\usepackage{amsmath} 
  
\DeclareMathOperator{\Fin}{finite}  
\DeclareMathOperator{\Diag}{diag}  
\usepackage{amsfonts}
\usepackage{color}   
\usepackage{hyperref}
\usepackage{multirow}
\usepackage{cancel}
\usepackage{slashed}
\usepackage{tensor}
\usepackage[table,xcdraw]{xcolor}

\title{\boldmath Two Point Functions in Defect CFTs}

\author{Christopher P.\ Herzog and Abhay Shrestha}
\affiliation{Department of Mathematics, King's College London,\\The Strand, London WC2R 2LS, UK}
\emailAdd{christopher.herzog@kcl.ac.uk, abhay.shrestha@kcl.ac.uk}

\abstract{
This paper is designed to be a practical tool for constructing and investigating two-point correlation functions
in defect conformal field theory, directly in physical space, between any two bulk primaries or between a bulk primary and a defect primary, with arbitrary spin. Although geometrically elegant and ultimately a more powerful approach, the embedding space formalism gets rather cumbersome when dealing with mixed symmetry tensors, especially in the projection to physical space. The results in this paper provide an alternative method for studying two-point correlation functions for a generic $d$-dimensional conformal field theory with a flat $p$-dimensional defect and $d-p=q$ co-dimensions. We tabulate some examples of 
correlation functions involving a conserved current, an energy momentum tensor and a Maxwell field strength, while analysing the constraints arising from conservation and the equations of motion. A method for obtaining bulk-to-defect correlators is also explained. Some explicit examples are considered: free scalar theory on $\mathbb{R}^p \times  ({\mathbb R}^q / {\mathbb Z}_2)$ and a free four dimensional Maxwell theory on a wedge. 
 }

\begin{document} 
\maketitle
\flushbottom

\newpage
\section{Introduction} \label{sec:intro}
Our interest is to improve fundamental understanding of defect conformal field theory (dCFT). 
Such an improvement
has a host of possible applications.  Conformal field theories with defects
and boundaries
describe physical systems at a phase transition,
for example water inside a container at the end-point of the liquid-gas critical line,
or an antiferromagnetic spin system with disorder at the N\'eel-dimer transition.   Topological defects may exist astrophysically, as relics from phase transitions in the early universe.
Thus an improved understanding of dCFT has experimental applications both in condensed matter and cosmology.
There are more fundamental theoretical reasons to pursue such a study as well, associated with
D-branes in string theory and twist defects for computing entanglement entropy in a quantum information context.
In this work, we study symmetry constraints on two-point correlation functions in dCFT.
\\

\noindent
Given substantial existing work
\cite{Billo,Lauria:2017wav,Kobayashi:2018okw,Guha,Marco}
using conformal symmetry to constrain correlation functions in dCFT,
one may legitimately ask why do more? 
Our answer is largely
personal, that in attempting to investigate some specific examples of dCFT correlation functions involving
tensors with mixed symmetries, for example a Maxwell field strength and a stress tensor, we found the embedding space formalism employed by refs.\ \cite{Billo,Lauria:2017wav,Kobayashi:2018okw,Guha,Marco} challenging to work with. 
The embedding space formalism requires an initial step of constructing a correlation function as a polynomial in embedding space,
and a second step, which is not always necessary, of projecting the polynomial to a tensorial object in physical space.  
\\

\noindent
The polynomials in embedding space are simpler objects than their tensorial physical space counter-parts. 
In many situations, it may be enough to work with these polynomials,
in which case the embedding space formalism remains a superior approach to ours. 
An old idea, it was originally developed in the context of CFT without boundaries or defects
\cite{Dirac,Weinberg,Costaa,Costa2}. 
The conformal group on $\mathbb{R}^d$ with $d>2$ is the orthogonal group $O(d+1,1)$ and its action is not linear. However, we know that $SO(d+1,1)$ acts naturally (and linearly) on $\mathbb{R}^{d+1,1}$. This fact can be exploited by embedding the spacetime into $\mathbb{R}^{d+1,1}$ and looking at the linearised action of the conformal group on this embedding space. The uplift simplifies the constraints on the $n$ point correlation functions arising from conformal symmetry.
\\

\noindent
Sometimes one would like access to the physical space correlation functions as well, however.  In our case, we wanted to use Feynman diagrams to investigate the effect of small interactions on the structure of correlation functions of free dCFT's.  As perturbation theory in CFT is typically performed in physical space, it is often more convenient to have a presentation of the conformal symmetry constraints on the physical space correlation functions.
(It is possible to lift the perturbative results to embedding space, contracting the tensor structures with polarization vectors.)
\\

\noindent
To refer to ``the'' embedding space formalism is already inaccurate, as there are at least two distinct variants of the formalism 
for mixed symmetry tensors, refs.\ \cite{Guha} and \cite{Marco}.  
  The ingredients that make up the polynomial are different in the two cases.
Ref.\ \cite{Guha} reproduces the antisymmetry of the correlation function through the use of Grassman variables.  
This approach has the advantage of producing simpler polynomials, but the disadvantage of making the projection
to physical space more difficult to implement.
Ref.\ \cite{Marco}, in contrast, encodes the antisymmetry of the correlators directly into bosonic building blocks for the polynomials.
While the projection to physical space is then as straightforward as in the fully symmetric case, the building blocks themselves
are more involved.
%
%
A further issue for us in both cases was that 
the projection procedure depends on whether the indices in the resultant tensors are tangent or perpendicular to the defect.
\\

\noindent
 The aim of this paper is to introduce a formalism where bulk correlation functions can be written directly in physical space 
 in a uniform way where parallel and perpendicular indices are treated on the same footing.
As such, we follow in the footsteps of McAvity and Osborn \cite{McAvityOsborn1} who developed a similar but simpler formalism for dealing with correlation functions of boundary CFT.  We hope that the formalism we present may be helpful to others. 
\\

\noindent
The work is organised as follows.
Section \ref{sec:conformalmaps} begins with a brief review of conformal maps to establish notation and conventions.
The second half of section \ref{sec:conformalmaps} presents our method for writing
down dCFT correlation functions directly in physical space, extending the boundary CFT formalism of ref.\ \cite{McAvityOsborn1}.
Section \ref{bulktwopointcorrelation} uses the formalism to present
several specific examples of two-point correlation functions involving
a scalar operator $\mathcal{O}$, a conserved current $J_\mu$, a stress tensor $T_{\mu\nu}$, and a Maxwell field (in four dimensions) $F_{\mu\nu}$. 
These tensorial objects are specified by symmetry constraints typically up to several functions of two invariant cross ratios.
In section \ref{sec:bulktodefectlimit}, we investigate the limit where one of the operators in these bulk two-point functions approaches the defect and hence derive the independent tensor structures required for writing any bulk-to-defect two point correlation function.  These are specified up to a handful of constants and we also present several examples.
Sections \ref{sec:free} looks at two specific examples of dCFT, a free scalar on
$\mathbb{R}^p \times(\mathbb{R}^q/\mathbb{Z}_2)$, a free Maxwell field on
${\mathbb R}^2 \times ({\mathbb R}^2 / {\mathbb Z}_N)$ for $N=2$ and 4 and also a free Maxwell field on ${\mathbb R} \times ({\mathbb R}^3 / {\mathbb Z}_2)$.
As supplemental material, 
we have provided a Mathematica notebook \cite{Mathematica} that defines some of the tensor structures we introduce and 
computes the $\langle F_{\mu\nu}(x) F_{\lambda \rho} (x') \rangle$ correlation function.

\section{Conformal Field Theory with a Flat Defect} \label{sec:conformalmaps}
Our principal interest is in formulating the conformal symmetry constraints on 
the  correlation functions of local operators in defect conformal field theory.
We begin with a brief review of conformal maps preserving the defect and their action on the local operators.
\\

\noindent
In the presence of a flat $p$-dimensional defect on a $d$-dimensional spacetime \cite{Billo}, the conformal symmetry is broken to a subgroup, $SO(p+1,1) \times SO(q),$ where $q=d-p$ is the codimension. Splitting spacetime into $\mathbb{R}^d=\mathbb{R}^p \times \mathbb{R}^q$, the defect is given by $\mathcal{D} = \{(\textbf{x},0)\in \mathcal{M} | \textbf{x} \in \mathbb{R}^p\}$. Any points away from the defect (where $y \neq 0$ with $(\textbf{x},y) \in \mathbb{R}^d$) we call the bulk space $\mathcal{B}$. Likewise, it is also convenient to split the index notation. We use Greek indices which run from $1,...,d$, Latin indices $a,b,c$ which run from $1,...,p$, and Latin indices $i,j,k$ which run from $p+1,...,d$ or from $1,...,q$. 
\\

\noindent
Now, we look at the conformal maps that preserve the defect. 
A conformal map $\phi$ preserves the defect if $\phi(p) \in \mathcal{D}$ for all $p\in \mathcal{D}$.
In the case of a flat defect defined above, the connected component of the conformal transformations on $\mathbb{R}^d$ which preserves the defect is given by, 
\begin{equation} \label{restrictedconf}
	\begin{aligned}
	\textbf{t}(\textbf{x},y) &= (\textbf{x}+\textbf{a},y), \\
	\textbf{r}_p(\textbf{x},y) &= (R_p \cdot \textbf{x},y), 
		\end{aligned} 
		\qquad 
		\begin{aligned}
	\textbf{r}_q(\textbf{x},y) &= (\textbf{x},R_q \cdot y), \\
\sigma(\textbf{x},y) &= (\sigma \textbf{x},\sigma y), 
		\end{aligned}
		\qquad
		\begin{aligned}
		\textbf{b}(\textbf{x},y) &= \left(\frac{\textbf{x}+\textbf{b}x^2}{\Omega(x)}, \frac{y}{\Omega(x)} \right).
		\end{aligned}
\end{equation}
where $x=(\textbf{x},y)$, $\textbf{a}\in \mathbb{R}^p$, $R_p \in SO(p)$, $R_q \in SO(q)$, $\sigma \in \mathbb{R}_{\neq 0}$, $\textbf{b}\in \mathbb{R}^p$ and $\Omega(x)=1+2 \textbf{b}\cdot \textbf{x} +\textbf{b}^2 x^2$. The full symmetry group is $SO(p+1,1)\times SO(q)$ and we will call a theory respecting this symmetry a defect CFT.
\\

\noindent
With a single bulk point, we are not able to form any conformal invariants (cross-ratios) and so the one point function is fixed up to a constant by conformal symmetry \cite{Billo}. Given two bulk points $x=(\textbf{x},y)$ and $x'=(\textbf{x}',y')$, we can construct two independent cross-ratios $\xi_a: \mathcal{B} \times \mathcal{B} \rightarrow \mathbb{R}$, under the defect conformal maps \eqref{restrictedconf},
\begin{equation} \label{crossratio}
\begin{split}
\xi_1 =\frac{s^2}{4 |y| |y'|}, 
\qquad
\xi_2 = \frac{y \cdot y'}{|y| |y'|}, 
\qquad 
s^2=(x-x')^2, 
\qquad
|y|=\sqrt{y_1^2+...+y_q^2}.
\end{split}
\end{equation}
(The defining property of a cross-ratio is $\xi_a(x,x')=\xi_a(\phi(x),\phi(x'))$.) 
Certain formulae are more simply expressed using particular rational combinations of $\xi_1$ and $\xi_2$. In particular, we will have occasion to use $u^2=\frac{\xi_1}{\xi_1+\xi_2}$ and $\xi_3=\frac{1-\xi_2^2}{\xi_2}$. 
\\

\noindent
For the special case $q=1$ we only have one independent cross-ratio since $\xi_2 \rightarrow 1$. Furthermore we have, $u^2 \rightarrow v^2=\frac{\xi_1}{\xi_1+1} $ and $\xi_3 \rightarrow 0$. 
The simplification in the case $q=1$ is part of a more general phenomena when we consider higher point functions.
As discussed in \cite{Lauria:2017wav,Guha}, given $n$ bulk points, we can construct the analog of $\xi_1$ and $\xi_2$
for any pair of these points, giving $n(n-1)$ cross ratios in general.  However, if $p$ or $q$ is too small compared to $n$, 
some of these cross ratios will not be independent, as happens when $n=2$ and $q=1$.  
More generally, there will be fewer independent cross ratios if $q < n$ or $p+2 < n$. For a lengthier and more detailed
discussion of these cross ratios, including when some of the operators live on the defect, see \cite{Lauria:2017wav}.

\subsection{$O(d)$ Vectors, Bi-vectors and Rank-2 Tensors} \label{subsec22}
\begin{table}[] 
	\centering
	\begin{tabular}{|l|l|l|l|l|l|l|}
		\hline \hline \hline
		\multicolumn{7}{|c|}{\textbf{Independent Tensor Structures}}                                                                                                                                                                                                                                 \\ \hline \hline \hline
		\multicolumn{1}{|c|}{\multirow{2}{*}{Co-Dimension}} & \multicolumn{1}{c|}{\multirow{2}{*}{Cross-Ratio}} & \multicolumn{2}{c|}{Vector}                                                        & \multicolumn{1}{c|}{\multirow{2}{*}{Bi-Vector}} & \multicolumn{2}{c|}{2-Tensor}  \\ \cline{3-4} \cline{6-7} 
		\multicolumn{1}{|c|}{}                              & \multicolumn{1}{c|}{}                              &\multicolumn{1}{c|}{at $x$}                                 &\multicolumn{1}{c|}{at $x'$}                                        & \multicolumn{1}{c|}{}                            &\multicolumn{1}{c|}{at $x$}        &\multicolumn{1}{c|}{at $x'$}         \\ \hline
		$q>2$                                                   & $\xi_1$, $\xi_2$                                   & $\Xi^{(1)}_{\mu}$, $\Xi^{(2)}_{\mu}$ & $\Xi'^{(1)}_{\alpha}$, $\Xi'^{(2)}_{\alpha}$ & $I_{\mu \alpha}$, $\mathcal{J}'_{\mu\alpha}$     &$\delta_{\mu \nu}$, $\mathcal{J}_{\mu \nu}$ &$\delta_{\alpha \beta}$, $\mathcal{J}''_{\alpha \beta}$ \\ \hline
		$q=2 $                                                & $\xi_1$, $\xi_2$                                   & $\Xi^{(1)}_{\mu}$, $\Xi^{(2)}_{\mu}$ & $\Xi'^{(1)}_{\alpha}$, $\Xi'^{(2)}_{\alpha}$ & $I_{\mu \alpha}$                                 & $\delta_{\mu \nu}$ & $\delta_{\alpha \beta}$ \\ \hline
		$q=1$                                                & $\xi_1$                                            & $\Xi^{(1)}_{\mu}$                    & $\Xi'^{(1)}_{\alpha}$                        & $I_{\mu \alpha}$                                 & $\delta_{\mu \nu}$ & $\delta_{\alpha \beta}$ \\ \hline \hline \hline
	\end{tabular}
	\caption{List of independent tensor structures used to construct two-point correlation function between any two bulk primaries with arbitrary spin. The definition of the cross-ratios, vectors, bi-vectors and rank-2 tensors can be found in \eqref{crossratio},  \eqref{confvectors}, \eqref{bivectors} and \eqref{rank2tensor} respectively. }
	\label{Table:tensorstructures}
\end{table}
The action of the defect conformal group \eqref{restrictedconf} on the correlation function must obey a Ward identity when the theory is a dCFT. The identity states that for any conformal transformation $\phi$ we have the equality,
\begin{equation} \label{WardIdentity}
\langle (\phi\cdot \mathcal{O}_I)(x) (\phi\cdot \mathcal{O}_J)(x') \rangle = \langle \mathcal{O}_I(x) \mathcal{O}_J(x') \rangle,
\end{equation}
where $\phi\cdot$ is the action of the conformal group on primary fields.  Recall that a primary field is defined by the transformation property, 
\[
(\phi\cdot \mathcal{O}_I)(\phi(x)) :=\Omega^{-\Delta}_{\phi}(x) \mathcal{G}^{\, J}_I[\mathcal{R}_{\phi}(x)] \mathcal{O}_J(x) \ ,
\]
 where $(\mathcal{R}_{\phi})^{\mu}_{\nu}(x) = \frac{(\partial_{\nu}\phi^{\mu})(x)}{\Omega_{\phi}(x)}$ and $\Delta$ is the scaling dimension of $\mathcal{O}_I$.  The $I,J$ are generalised  indices that indicate the representation of $O(d)$ under which $\mathcal{O}$ transforms. Lastly, $\mathcal{G}$ is a matrix acting on the representation space of $O(d)$. 
 \\
 
 \noindent
 Hence, \eqref{WardIdentity} provides constraints on the correlation function arising from the defect conformal group. In table \ref{Table:tensorstructures}, we list a set of independent tensor structures that can be used to construct correlation functions satisfying the Ward identity.
\\

\noindent
Using the cross-ratios \eqref{crossratio}, we can define two structures $\Xi^{(1)}$ and $\Xi^{(2)}$ which enable us to satisfy the Ward indentity for primary vectors. The structures $\Xi^{(a)}$ can be viewed as $O(d)$ vectors at a point $x$; they transform as $\Xi^{(a)}_{\mu} \rightarrow (\mathcal{R}_{\phi})^{\nu}_{\mu}(x) \Xi^{(a)}_{\nu}$ under the defect conformal group. 
Along with $\Xi^{(a)}$, we can define two vectors $\Xi'^{(n)}$ which transforms as an $O(d)$ vector at $x'$. A choice of these are given explicitly in Cartesian coordinates as
\begin{equation} \label{confvectors}
\begin{aligned}
\Xi^{(1)}_{\mu}(x,x') &= \frac{|y|}{\xi_1} \frac{\partial \xi_1}{\partial x^{\mu}} = \frac{2|y|}{s^2}s_{\mu} -n_{\mu}, \\
\Xi^{(2)}_{\mu}(x,x') &=\frac{|y|}{\xi_2} \frac{\partial \xi_2}{\partial x^{\mu}} = \frac{n'_{\mu}}{\xi_2}-n_{\mu}, 
\end{aligned}
\qquad
\begin{aligned}
\Xi'^{(1)}_{\mu}(x,x') &= \frac{|y'|}{\xi_1} \frac{\partial \xi_1}{\partial x'^{\mu}} = -\frac{2|y'|}{s^2}s_{\mu} -n'_{\mu}, \\
\Xi'^{(2)}_{\mu}(x,x') &= \frac{|y'|}{\xi_2} \frac{\partial \xi_2}{\partial x'^{\mu}} = \frac{n_{\mu}}{\xi_2}-n'_{\mu}, 
\end{aligned}
\end{equation}
where,
\begin{equation} \label{normalvec}
	\begin{aligned}
	n_{\mu} = \begin{cases} 
	0 & \mu=a,  \\
	\frac{y_k}{|y|} & \mu=k,
	\end{cases} \\
	\end{aligned}
	\qquad
	\begin{aligned}
		n'_{\mu} = \begin{cases} 
		0 & \mu=a,  \\
		\frac{y'_k}{|y'|} & \mu=k,
		\end{cases}
	\end{aligned}
\end{equation}
and $s_{\mu}=x_{\mu}-x'_{\mu}$. With $\Xi^{(a)}$ and $\Xi'^{(n)}$ in hand, 
we can construct bi-vectors, which 
 transform as an $O(d)$ vector at $x$ and $x'$, by taking the product $\Xi^{(a)} \Xi'^{(n)}$. 
Similarly, we can form rank-2 $O(d)$ tensors at $x$ by taking the product $\Xi^{(a)}\Xi^{(b)}$ and ones at $x'$ by taking the product $\Xi'^{(n)} \Xi'^{(m)}$. 
\\

\noindent
 We can also take further derivatives of the vectors \eqref{confvectors}. Since $\mathcal{R}_{\phi}$ depends on $x$ or $x'$, we can only take the derivative w.r.t.~$x$ of $\Xi'$ and w.r.t.~$x'$ of $\Xi$.  Two derivatives w.r.t.~the same point results in an object which will not transform correctly. These derivatives yield two additional bi-vectors:
\begin{subequations} \label{bivectors}
	\begin{align}
   I_{\mu \nu}(x,x') &= -2\xi_1|y| \frac{\partial}{\partial x^{\mu}} \Xi'^{(1)}_{\nu}=-2\xi_1|y'| \frac{\partial}{\partial x'^{\mu}} \Xi^{(1)}_{\nu}, \\
   \mathcal{J}'_{\mu \nu}(x,x') &= \xi_2 |y| \frac{\partial}{\partial x^{\mu}} \Xi'^{(2)}_{\nu} = \xi_2 |y'| \frac{\partial}{\partial x'^{\nu}} \Xi^{(2)}_{\mu},
	\end{align}
\end{subequations}
where $I_{\mu \nu} = \delta_{\mu \nu} - \frac{2s_{\mu}s_{\nu}}{s^2}$, is the rotation matrix corresponding to the inversion map \cite{McAvityOsborn2}, $I_{\mu \nu}=(\mathcal{R}_{\rm inv})_{\mu \nu}(x-x')$ and
\begin{equation} \label{bivectorJ}
\mathcal{J}'_{\mu \nu}(x,x') = \begin{cases} 
\delta_{ij}-\frac{y_j y'_i}{y\cdot y'} & \mu=i, \nu=j \\
0 & {\rm otherwise}.
\end{cases}
\end{equation}
\\

\noindent
The last two independent rank-2 tensors come from contracting two bivectors over the indices which transform at the same point. The contractions between $\mathcal{J}'$ and $I$ or $\mathcal{J}'$ gives a new rank-2 tensor at $x$ and 
a second at $x'$:
\begin{equation} \label{rank2tensor}
	\begin{split}
	\mathcal{J}_{\mu \nu}(x) = \begin{cases} 
	\delta_{ij}-n_in_j & \mu=i, \nu=j \\
	0 & {\rm otherwise}
	\end{cases},
	\quad
	\mathcal{J}''_{\mu\nu}(x') =  \begin{cases} 
	\delta_{ij}-n'_in'_j & \mu=i, \nu=j \\
	0 & {\rm otherwise}
	\end{cases}.
	\end{split}
\end{equation}
The number of primes on $\mathcal{J}$ indicates how many times the point $x'$ is implicated in its transformation
properties: no primes means a rank-2 tensor at $x$, one prime a bivector at $x$ and $x'$, and two primes a rank-2 tensor at $x'$.
\\

\noindent
We now demonstrate by explicit computation that the set of tensor structures in table \ref{Table:tensorstructures}
is closed under index contraction. We consider only contractions of indices associated with the point $x$ since the same relations will hold for $x'$ 
under the replacement $\Xi \rightarrow \Xi'$, $\mathcal{J} \rightarrow \mathcal{J}''$.   The contractions involving vectors
and rank-2 tensors at $x$ are
\begin{equation} \label{contractions1}
\begin{aligned}
\Xi^{(1)}_{\mu} \Xi^{(1) \mu} &= \frac{1}{u^2}, \\
\Xi^{(2)}_{\mu} \Xi^{(2) \mu} &= \frac{\xi_3}{\xi_2}, \\
\Xi^{(1)}_{\mu} \Xi^{(2) \mu} &= -\frac{1}{2} \frac{\xi_3}{\xi_1}, \\
\end{aligned}
\qquad
\begin{aligned}
\mathcal{J}_{\mu \nu} \Xi^{(1) \nu} &= -\frac{\xi_2}{2\xi_1} \Xi^{(2)}_{\mu},\\
\mathcal{J}_{\mu \nu} \Xi^{(2) \nu} &= \Xi^{(2)}_{\mu}, \\
\mathcal{J}_{\mu \alpha} \mathcal{J}^{\alpha}_{\nu} &= \mathcal{J}_{\mu \nu}, \\
\mathcal{J}^{\mu}_{\mu} &=q-1.
\end{aligned}
\end{equation}
The contractions involving the bi-vectors are,
\begin{equation} \label{contractions2}
\begin{aligned}
I_{\mu \nu} \Xi^{(1) \nu} &=\mathcal{X}'_{\mu}, \\
I_{\mu \nu} \Xi^{(2) \nu} &= -(\xi_3\Xi'^{(1)}_{\mu} + \xi_2 \Xi'^{(2)}_{\mu}), \\
\mathcal{J}'_{\mu \nu} \Xi^{(1) \mu} &= \frac{1}{2\xi_1} \Xi'^{(2)}_{\nu}, \\
\mathcal{J}'_{\mu \nu} \Xi^{(2) \mu} &= -\frac{1}{\xi_2}\Xi'^{(2)}_{\nu}, \\
\end{aligned}
\qquad
\begin{aligned}
\mathcal{J}'_{\mu \alpha} (\mathcal{J}')_{\nu}^{\; \; \alpha} &= \mathcal{J}_{\mu \nu} + \Xi^{(2)}_{\mu} \Xi^{(2)}_{\nu}, \\
\mathcal{J}'_{\mu\alpha}I^{\alpha}_{\nu} &= \mathcal{J}_{\mu \nu} + \Xi^{(2)}_{\mu} \Xi^{(1)}_{\nu}, \\
\mathcal{J}_{\mu \nu} I^{\nu}_{\alpha} &=\mathcal{J}'_{\mu \alpha}-\Xi^{(2)}_{\mu}\mathcal{X}'_{\alpha},\\
\mathcal{J}_{\mu}^{\nu} \mathcal{J}'_{\nu \alpha} &= \mathcal{J}'_{\mu \alpha}, \\
I_{\mu \alpha} I^{\alpha}_{\mu} &= \delta_{\mu \nu}.
\end{aligned}
\end{equation}
where $\mathcal{X}'_{\mu}:=\xi_2(\Xi'^{(1)}_{\mu}-\Xi'^{(2)}_{\mu})$. 
\\

\noindent
The structures in table \ref{Table:tensorstructures} are independent in the sense that they cannot be written as a product of lower rank tensor structures.  In this language, $\Xi^{(n)} \Xi'^{(m)}$ is not an independent bi-vector even though it is a necessary ingredient in constructing the correlation function of two vector operators.
The independent structures are closed under contraction, as seen from \eqref{contractions1} and \eqref{contractions2}.
No new ones can be formed through derivatives or contractions.  \\

\noindent
For a parity even theory, we expect that this set of structures is 
suficient to construct the correlation function between any two bulk operators of arbitrary spin.\footnote{%
For parity odd theories, we must add some Levi-Civita tensors $\epsilon^{\mu_1 \cdots \mu_d}$, $\epsilon^{a_1 \cdots a_p}$ and $\epsilon^{i_1 \cdots i_q}$ to the construction.  
} 
For bulk operators with $I_1$ and $I_2$ indices, we construct all possible terms with $I_1$ and $I_2$ indices
using the independent structures in table \ref{Table:tensorstructures}.  We then symmetrize or antisymmetrize over the indices and remove traces, as required in order to obtain an object with the right transformation properties under the 
$O(d) \times O(d)$ group acting on the two operators.\\

\noindent
Indeed, not counting the Kronecker delta function used to remove traces, 
the number eight of independent structures here is the same as the number of structures used to construct
correlation function polynomials in embedding space in \cite{Billo}.  Moreover, one can see schematically 
that the monomial
building blocks in embedding space project down to elements which have the same spin as our structures in real space.
Therefore, the number of independent terms in any correlation function that we construct must match ref.\ \cite{Billo}.
We only consider up to rank-2 tensors in this work.   However, counting the total number of structures in selected correlators such as $\langle S_{\mu \nu \rho} \mathcal{O} \rangle$, $\langle S_{\mu \nu \rho} V_{\alpha} \rangle$, $\langle S_{\mu \nu \rho \lambda} \mathcal{O} \rangle$, for symmetric, traceless $S$, we can match the number obtained from embedding 
space.\footnote{%
Eq.\ (3.19) in of ref.\  \cite{Billo} is missing a couple of factors.  The correct version should be (pers.\ comm.\ E.~Lauria)
\[
\sum_{k=0}^{{\rm min}(J_1, J_2)} \prod_{j=1}^2 \left(J_j - k + 1 - \left\lfloor \frac{J_j -k}{2} \right\rfloor \right)
\left( \left\lfloor \frac{J_j-k}{2} \right\rfloor + 1 \right) \ .
\]
} 
We also match the number, six, of independent structures in the correlator $\langle F_{\mu\nu} V_\lambda \rangle$ for an antisymmetric
operator $F_{\mu\nu}$ (pers.\ comm.\ M.~Meineri).
\paragraph{$\textbf{q}=\textbf{1},\textbf{2}$:} When $q=1$, we reduce to bCFT. Here $\Xi^{(2)}=\Xi'^{(2)}=\mathcal{J}=\mathcal{J}'=\mathcal{J}''=0$. Similarly, $q=2$ is also a special case because we find that $\mathcal{J}$, $\mathcal{J}'$ and $\mathcal{J}''$ are not independent. In this case the following identities hold,
\begin{equation}
	\begin{split} \label{Jprimenotindp}
	\mathcal{J}'_{\mu \nu} =-\frac{1}{\xi_3}\Xi^{(2)}_{\mu}\Xi'^{(2)}_{\nu}, 
	\qquad
	\mathcal{J}_{\mu \nu} = \frac{\xi_2}{\xi_3}\Xi^{(2)}_{\mu}\Xi^{(2)}_{\nu}, 
	\qquad
	\mathcal{J}''_{\mu \nu} = \frac{\xi_2}{\xi_3}\Xi'^{(2)}_{\mu}\Xi'^{(2)}_{\nu}.
	\end{split}
\end{equation}
\subsection{Comment about one and higher point functions} \label{sec:higherpoint}
\noindent
The main purpose of this work is to investigate two point functions, but we would like to make a couple of remarks 
about one point and higher point functions before passing to the main order of business.
Curiously, although we found the tensor ${\mathcal J}_{\mu\nu}$ through the existence of cross ratios, this structure
exists independently of them, and is important for allowing nonzero one-point functions in dCFT.
If we have an operator ${\mathcal O}_I$ in a representation of $O(d)$ such that we can also construct something out 
of the ${\mathcal J}_{\mu\nu}$ and $\delta_{\mu\nu}$ structures in the same representation, then ${\mathcal O}_I$ is allowed to have a nonzero one point function.  Importantly, ${\mathcal J}_{\mu\nu}$ does not exist for $q=1$ which forbids anything
except for scalars developing a nonzero expectation value in bCFT \cite{McAvityOsborn1}. 
More generally in dCFT, 
we see that vectors and anti-symmetric two-forms are also forbidden from having a nonzero expectation value.
\\

\noindent
Having gone through the exercise of constructing two-point correlation functions, the procedure in broad outline is clear for $n$-point functions as well. 
Given $n$ bulk points, for each pair $(x_r,x_s)$ we can construct $\xi_1$ and $\xi_2$ type cross-ratios, 
calling them $\xi_1^{(r,s)}$ and $\xi_2^{(r,s)}$ respectively. 
Since a total of $n(n-1)/2$ unique pairs can be formed, we have naively $n(n-1)$ independent cross ratios.
 For $p$ and $q$ to small compared to $n$, some of these won't be independent (see the discussion immediately preceeding Sec.\ \ref{subsec22}). 
\\

\noindent
Given our set of independent cross ratios, we can then construct tensor structures analogous to those in 
figure \ref{Table:tensorstructures}.  Fixing a point $x_r$, we can form $(n-1)$ distinct pairs involving $x_r$. Taking derivatives with respect to $x_r$ of both $\xi_1^{(r,s)}$ and $\xi_2^{(r,s)}$ gives us
 $2(n-1)$ independent vectors at the given point $x_r$. Repeating this procedure 
 for all the $n$ points gives a total of $2n(n-1)$ vectors, 
$\Xi_{\mu}^{(1)(r,s)}$ and $\Xi_{\mu}^{(2)(r,s)}$. 
%
%
By further taking the derivative of $\Xi^{(1)(r,s)}$ and $\Xi^{(2)(r,s)}$ with respect to $x_s$,
 we can form a bi-vector of type $I$ and type $\mathcal{J}'$.  Let us call them $I_{\mu \nu}^{(r,s)}$ and $\mathcal{J}_{\mu \nu}'^{(r,s)}$ respectively.  
 Consequently, at each pair we have 2 independent bi-vectors and hence a total of $n(n-1)/2$ bi-vectors of type $I$ and $\mathcal{J}'$ each. Finally, for each point $x_r$ we can construct the independent rank-2 tensor $\mathcal{J}$, namely $\mathcal{J}^{(r)}_{\mu \nu}$. 
 \\
 
 \noindent
 We then assemble from these constituents objects with the correct $O(d)$ transformation properties -- by appropriately
antisymmetrizing, symmetrizing, and removing traces.
 Although we do not check the closure of this set of independent tensor structures under contraction (as done for the two-point case in \eqref{contractions1} and \eqref{contractions2}), we do find a correspondence with the embedding space results in \cite{Guha}. In particular, the number of vectors matches their $K_{ab}^{(i)}$ and $\bar{K}_{ab}^{(i)}$, the number of bi-vectors matches their $S_{ab}^{(i,j)}$ and $\bar{S}_{ab}^{(i,j)}$ and finally the number of rank-2 tensors matches their $H_a^{(i,j)}$.

\section{Bulk-Bulk Two Point Functions} \label{bulktwopointcorrelation}
\begin{table}[h]
	\centering
	\begin{tabular}{|c|c|c|c|c|c|c|}
		\hline \hline \hline
		\multirow{2}{*}{Correlators}         & \multicolumn{3}{c|}{\# of functions}                                                       & \multicolumn{3}{c|}{Conservation}                                                          \\ \cline{2-7} 
		& $q > 2$                      & $ q = 2$                     & $q = 1$                      & $q > 2$                      & $q = 2$                      & $q = 1$                      \\ \hline
		$\langle J_{\mu} \mathcal{O} \rangle$      & 2                            & 2                            & 1                            & 1                            & 1                            & 1                            \\ \hline
		$\langle J_{\mu} J_{\nu} \rangle$      & 5                            & 4                            & 2                            & 2                            & 2                            & 1                            \\ \hline
		$\langle T_{\mu \nu} \mathcal{O} \rangle$      & 4                            & 4                            & 1                            & 2                            & 2                            & 1                            \\ \hline
		$\langle T_{\mu \nu} J_{\alpha}\rangle$      & 12                           & 8                            & 2                            & 10                           & 8                            & 3                            \\ \hline
		$\langle T_{\mu \nu} T_{\alpha \beta} \rangle$      & 19                           & 10                           & 3                            & 12                           & 8                            & 2                            \\ \hline \hline \hline
		\multirow{2}{*}{Correlators, $d=4$} & \multicolumn{3}{c|}{\# of functions}                                                       & \multicolumn{3}{c|}{EoM}                                                                   \\ \cline{2-7} 
		& \multicolumn{1}{l|}{$q = 3$} & \multicolumn{1}{l|}{$q = 2$} & \multicolumn{1}{l|}{$q = 1$} & \multicolumn{1}{l|}{$q = 3$} & \multicolumn{1}{l|}{$q = 2$} & \multicolumn{1}{l|}{$q = 1$} \\ \hline
		$\langle F_{\mu \nu} \mathcal{O} \rangle$      & 1                            & 1                            & 0                            & 2                            & 2                            & 0                            \\ \hline
		$\langle F_{\mu \nu} F_{\alpha \beta} \rangle$      & 10                           & 5                            & 2                            &6                         & 4                            & 1                            \\ \hline \hline \hline
	\end{tabular}
	\caption{This table contains a list of bulk-bulk correlators which we consider in section \ref{bulktwopointcorrelation} and the number of independent structures appearing in the correlators, dependent on the co-dimension $q$. It also denotes the number of PDE constraints arising from conservation of current $\partial^{\mu} J_{\mu}=0$ and stress tensor $\partial^{\mu}T_{\mu \nu}=0$, everywhere in the bulk. Correlators involving a Maxwell field strength $F_{\mu \nu}$ are also considered, specifically when $d=4$, and we list the number of PDE constraints arising from the bulk equation of motion $\partial^{\mu}F_{\mu \nu}=0$.}
	\label{bulkcorrelatorlist}  
\end{table}
\noindent
As a warm-up, the two point correlator between two bulk scalar primaries of dimension $\Delta$ and $\Delta'$ is well known to have the form\footnote{Note we can use \eqref{crossratio} to write $|y|=\frac{s^2}{4|y'|\xi_1}$.}
\begin{equation} \label{bulkscalartwopoint}
\langle \mathcal{O}(x) \mathcal{O'}(x') \rangle = \frac{|y'|^{\Delta-\Delta'}}{s^{2\Delta}} f(\xi_1, \xi_2),
\end{equation}
where $f(\xi_1, \xi_2)$ is an arbitrary function of the cross-ratios. In what follows, we will investigate correlation functions involving a scalar field ${\mathcal O}$, a conserved current $J^\mu$, and stress tensor $T^{\mu\nu}$ and (in the particular case of four dimensions) a Maxwell field strength $F^{\mu\nu}$. Also, when counting the number of PDE constraints arising from conservation or equations of motion (see table \ref{bulkcorrelatorlist}), we use the argument that taking the divergence reduces the spin of the correlator by 1 and hence the number of independent structures present in the resulting correlator is equal to the number of PDE constraints.  For example taking the divergence of $\langle T_{\mu \nu} \mathcal{O} \rangle$ we see the resulting correlator is of the form $\langle V_{\nu} \mathcal{O} \rangle$ and hence 2 independent PDE constraints are expected.
\subsection{$\langle J \mathcal{O} \rangle$}
The two point correlator between a bulk vector operator $V_\mu$ of dimension $\Delta$ and a scalar primary  
${\mathcal O}$ of dimension $\Delta'$ is fixed up to two functions of two cross ratios:
\begin{equation} \label{VObulk1}
\langle V_{\mu}(x) \mathcal{O}(x') \rangle = \frac{1}{|y|^{\Delta} |y'|^{\Delta'}} \left(f_1(\xi_1, \xi_2) \Xi^{(1)}_{\mu} + f_2 (\xi_1, \xi_2) \Xi^{(2)}_{\mu}  
\right) \ . 
\end{equation}
In the special case that $V^\mu = J^\mu$ is a conserved current (with dimension $\Delta = d-1$), we have the constraint
$\partial_{\mu} \langle J^{\mu}(x) \mathcal{O}(x') \rangle =0$, satisfied everywhere in the bulk.
Current conservation leads to the following relation between the functions $f_1$ and $f_2$:
$$
(2p \xi_1 + (d-2) \xi_2) f_1+\frac{2\xi_1^2}{u^2}f_1^{(1,0)}-\xi_2\xi_3f_1^{(0,1)} 
$$
\begin{equation} \label{JOPDE}
+ 2 \xi_1 \left( 2- q - \frac{1}{\xi_2^2} \right) f_2
 -\xi_1\xi_3 \left(f_2^{(1,0)}-2f_2^{(0,1)} \right) =0.
\end{equation}
where the identities \eqref{JOidentities1} were used in the derivation. 
\paragraph{$\textbf{q}=\textbf{1}$:} 
In the codimension one case, 
the structure $\Xi^{(2)}$ is absent and 
the cross ratios $\xi_2 \rightarrow 1$ and $\xi_3 \rightarrow 0$ degenerate.  The constraint \eqref{JOPDE} reduces to,
\begin{equation*}
((d-2) + 2 (d-1)\xi_1) f_1 + 2 \xi_1 (1+\xi_1) \frac{df_1}{d\xi_1} = 0 
\end{equation*}
which has the simple solution
\[
f_1(\xi_1) = c \, \xi_1^{1-\frac{d}{2}} (1+\xi_1)^{-\frac{d}{2}} = c \, \xi_1^{1-d} \, v^d \ ,
\]
where $c$ is an integration constant.  This result matches the bCFT result in \cite{McAvityOsborn2} with appropriate rescaling.
\subsection{$\langle JJ \rangle$}
The two point correlator between two identical vector fields of dimension $\Delta$
 depends on five arbitrary functions of two cross ratios:
\begin{equation} \label{VVid}
\begin{split}
\langle V_{\mu}(x) V_{\nu}(x') \rangle = \frac{1}{s^{2\Delta}} \bigg(&f_1\Xi^{(1)}_{\mu} \Xi'^{(1)}_{\nu} + f_2\Xi^{(2)}_{\mu} \Xi'^{(2)}_{\nu} + f_3(\Xi^{(1)}_{\mu} \Xi'^{(2)}_{\nu}+ \Xi^{(2)}_{\mu} \Xi'^{(1)}_{\nu})\\
\qquad
&+f_4 I_{\mu \nu} + f_5 \mathcal{J}'_{\mu \nu} \bigg),
\end{split}
\end{equation}
The  tensor structure of $f_3$ appears in a symmetric combination
because the operators in the correlation function are assumed to be identical.
(Conversely, 
if the operators were distinct, 
the coefficients of $\Xi^{(1)}_{\mu} \Xi'^{(2)}_{\nu}$ and $\Xi^{(2)}_{\mu} \Xi'^{(1)}_{\nu}$
should be independent.) 
\\

\noindent
Reflection positivity
places bounds on the bulk-bulk functions appearing in the correlator between identical operators. 
The reflection plane is taken to be a hypersurface that intersects the defect at right angles, which
fixes $\xi_2 = 1$.   
The positivity demands that,
\begin{equation} \label{boundonJJ}
\begin{split}
f_4(\xi_1,1) \geq 0, 
\qquad
(f_4+f_5)(\xi_1,1) \geq 0,
\qquad 
\left(f_4 + \frac{f_1}{u^2} \right)(\xi_1,1) \geq0.
\end{split}
\end{equation}
If $V^\mu = J^\mu$ is a conserved current, then $\Delta = d-1$ and 
$\partial_\mu \langle J^\mu(x) J^\nu(x') \rangle = 0$.  
This divergence $\partial_\mu \langle J^\mu(x) J^\nu(x') \rangle$ is the correlation function of a scalar with a vector, and we just saw that it depends generically on 
two functions of two cross ratios.  Thus, conservation will lead to two constraints on the five $f_i$.
Using the identities \eqref{JOidentities1} and \eqref{JJidentities}, we find
\begin{equation} \label{JJPDE1}
\begin{aligned}
\frac{2\xi_1^2}{u^2} f_1^{(1,0)}&-\xi_2\xi_3f_1^{(0,1)}- \left(\xi_2(d+1)+2\xi_1(q-1) \right)f_1 +\xi_1\xi_3\left(2f_3^{(0,1)}-f_3^{(1,0)} \right) \\
+&\left(\xi_3d -2\xi_1\left(\frac{\xi_3}{\xi_2}+(q-1)\right)\right)f_3 +2\xi_1^2\xi_2f_4^{(1,0)}-2\xi_1\xi_2\xi_3f_4^{(0,1)}=0,
\end{aligned}
\end{equation}
and
\begin{equation} \label{JJPDE2}
\begin{aligned}
\xi_2 f_1 &+ \xi_1\xi_3\left(2f_2^{(0,1)}-f_2^{(1,0)}\right) + \left(\xi_3(d-1)-2\xi_1\left(\frac{\xi_3}{\xi_2}+\frac{1}{\xi_2^2}+(q-1) \right)\right)f_2 \\
&+\frac{2\xi_1^2}{u^2}f_3^{(1,0)}-\xi_2\xi_3f_3^{(0,1)}-\left(\xi_2(d-1)+2\xi_1(q-1)-\frac{1}{\xi_2}\right)f_3-2\xi_1^2\xi_2f_4^{(1,0)} \\
&-2\xi_1\xi_2^2f_4^{(0,1)}+\xi_1\left(f_5^{(1,0)}-2f_5^{(0,1)}\right)+\left(\frac{2\xi_1}{\xi_2}-(d-1)\right)f_5 =0.
\end{aligned}
\end{equation}
\paragraph{$\textbf{q}=\textbf{1}$:} Here we only have one PDE because (\ref{JJPDE2}) was obtained by setting the coefficient of $\Xi'^{(2)}$ to zero, which does not exist when $q=1$. The PDE constraint for $q=1$ is,
\begin{equation*}
2\xi_1^2\left(\frac{1}{v^2}\frac{df_1}{d\xi_1}+\frac{df_4}{d\xi_1}\right) = (d+1)f_1.
\end{equation*}
Notice that this only involves $f_1$ and $f_4$ since the other tensor structures are zero when $q=1$. The PDE constraint matches the bCFT result in \cite{McAvityOsborn2} under appropriate identification. 
\paragraph{$\textbf{q}=\textbf{2}$:} For this case we take $f_5 \rightarrow 0$ since $\mathcal{J}'$ is not independent \eqref{Jprimenotindp}. 

\subsection{$\langle T \mathcal{O} \rangle$}
The two point correlator of a symmetric rank-2 tensor $S$ with dimension $\Delta$ and a scalar of dimension $\Delta'$ is given by,\footnote{%
We use round brackets on indices to denote symmetrisation and square brackets for 
antisymmetrisation (including a factor of $\frac{1}{n!})$. 
We use $|$ to separate indices that are being (anti)symmetrised when not next to each other. 
}
\begin{equation}
\langle S_{\mu \nu}(x) \mathcal{O}(x') \rangle = \frac{|y'|^{\Delta-\Delta'}}{s^{2\Delta}} \bigg(f_1\Xi^{(1)}_{\mu}\Xi^{(1)}_{\nu} + f_2\Xi^{(2)}_{\mu}\Xi^{(2)}_{\nu} + f_3\Xi^{(1)}_{(\mu}\Xi^{(2)}_{\nu)} + f_4 \mathcal{J}_{\mu \nu} + f_5 \delta_{\mu \nu} \bigg).
\end{equation}
We need the correlator to be traceless to consider the energy momentum tensor. Taking the trace of the above equation, we obtain a constraint on $f_5$,
\begin{equation}
f_5=- \left(\frac{f_1}{u^2 d} +\frac{f_2 \xi_3}{\xi_2 d} -\frac{f_3 \xi_3}{2\xi_1 d} + \frac{f_4(q-1)}{d} \right).
\end{equation}
Now we can write the two point correlator between the energy momentum tensor and a scalar,
\begin{equation}
\begin{split}
\langle T_{\mu \nu}(x) \mathcal{O}(x') \rangle = \frac{|y'|^{d-\Delta'}}{s^{2d}} &\bigg[f_1\left(\Xi^{(1)}_{\mu}\Xi^{(1)}_{\nu}-\frac{\delta_{\mu \nu}}{u^2 d}\right) + f_2\left(\Xi^{(2)}_{\mu}\Xi^{(2)}_{\nu} - \frac{\delta_{\mu \nu} \xi_3}{\xi_2 d}\right) \\
+&f_3\left(\Xi^{(1)}_{(\mu}\Xi^{(2)}_{\nu)} + \frac{\delta_{\mu \nu} \xi_3}{2\xi_1 d} \right) + f_4\left(\mathcal{J}_{\mu\nu}-\frac{q-1}{d}\delta_{\mu \nu} \right) \bigg].
\end{split}
\end{equation}
This gives a total of 4 structures matching the number obtained from the embedding space theory \cite{Billo}. Imposing conservation of $T$ and using the identities \eqref{TOidentities}, we obtain two PDE constraints on $f_1,...,f_4$,
\begin{equation} \label{firstTO}
\begin{split}
\left( 2\xi_1(1-q) -\frac{\xi_2}{d}(d+2)(d-1) \right)f_1 &+ \frac{2\xi_1^2}{u^2d}(d-1)f_1^{(1,0)} -\xi_2\xi_3 f_1^{(0,1)} + \frac{2\xi_1\xi_3}{\xi_2}f_2 \\
- \frac{2\xi_1^2\xi_3}{\xi_2d}f_2^{(1,0)} &+ \left(\frac{\xi_3}{2d}(d+1)(d-2) -\xi_1 \left(\frac{\xi_3}{\xi_2}+q-1 \right) \right)f_3  \\
+\xi_1\xi_3 \left(\frac{2-d}{2d}f_3^{(1,0)} + f_3^{(0,1)} \right) &+2(q-1)\xi_1 f_4 -2\xi_1^2\left(\frac{q-1}{d} \right)f_4^{(1,0)} = 0,
\end{split}
\end{equation}
and,
\begin{equation} \label{secondTO}
\begin{aligned}
&\frac{\xi_2}{d}(d-2)f_1 - \frac{2\xi_1 \xi_2}{u^2d}f_1^{(0,1)}  +\left(\xi_3d -4\xi_1 \left(\frac{\xi_3}{\xi_2} - \frac{1}{\xi_2^2d} + \frac{q}{2} \right)   \right)f_2 \\
&+\xi_1\xi_3 \left(\frac{2}{d}(d-1)f_2^{(0,1)} -f_2^{(1,0)} \right) + \left(\frac{\xi_3}{2} -q\xi_1 -\frac{\xi_2}{d}\left(1+\frac{d^2}{2} +\frac{1}{\xi_2^2} \right) \right)f_3 \\
&+\frac{\xi_1^2}{u^2}f_3^{(1,0)} + \frac{\xi_2\xi_3}{2d}(2-d)f_3^{(0,1)} + \xi_2d f_4 -\xi_1\xi_2 \left(f_4^{(1,0)} + 2\left(\frac{q-1}{d}-1 \right)f_4^{(0,1)} \right) = 0.
\end{aligned}
\end{equation}
\paragraph{$\textbf{q}=\textbf{1}$:} For this case $f_2$,...,$f_4$ vanish, $\xi_2 \rightarrow 1$ and $\xi_3 \rightarrow 0$, while the second PDE \eqref{secondTO} (obtained by setting coefficients of $\Xi^{(2)}$ to zero) does not exist. So, the constraint reduces to,
\begin{equation*}
\frac{2\xi_1^2}{v^2} \frac{df_1}{d\xi_1} = (d+2)f_1.
\end{equation*}
This is simply solved to give,
\begin{equation*}
f_1(\xi_1)=c \left(\frac{\xi_1}{\xi_1+1} \right)^{1+\frac{d}{2}} \implies f_1(v)=c v^{d+2},
\end{equation*}
which matches the result from \cite{McAvityOsborn2} with the appropriate rescaling. 

\paragraph{$\textbf{q}=\textbf{2}$:} Here, $\mathcal{J}$ is not independent and so we set $f_4=0$. 

\subsection{$\langle TJ \rangle$}
The two point correlator between a symmetric rank-2 tensor $S$ of dimension $\Delta$ and a vector $V$ of dimension $\Delta'$ is given by,
$$
\langle S_{\mu \nu}(x) V_{\alpha}(x') \rangle = \frac{|y'|^{\Delta-\Delta'}}{s^{2\Delta}} \bigg(g_n \mathcal{J}_{\mu \nu} \Xi'^{(n)}_{\alpha}+h_n\delta_{\mu\nu}\Xi'^{(n)}_{\alpha} + F_n\Xi^{(n)}_{(\mu}I^{\phantom{(n)}}_{\nu) \alpha} +G_n \Xi^{(n)}_{(\mu} \mathcal{J}'_{\nu) \alpha}
$$
\begin{equation} \label{TJcorrelator}
+\sum_{n \geq m,r}f_{mnr}\Xi^{(m)}_{(\mu}\Xi^{(n)}_{\nu)} \Xi'^{(r)}_{\alpha} \bigg),
\end{equation}
where we restrict the sum over $n$ to avoid double counting and so $f_{mnr}$ has 6 independent components. There are 14 structures here, but tracelessness of $S_{\mu\nu}$ removes two, constraining $h_1$ and $h_2$,
\begin{subequations}
	\begin{align} \label{TJh1}
	h_1 &=-\frac{1}{d} \left(g_1(q-1) + \frac{f_{111}}{u^2}-\frac{f_{121}\xi_3}{2\xi_1} +\frac{f_{221}\xi_3}{\xi_2}+ F_1 \xi_2 - F_2 \xi_3 \right), \\ \label{TJh2}
	h_2 &=-\frac{1}{d}\left(g_2(q-1)+\frac{f_{112}}{u^2}-\frac{f_{122}\xi_3}{2\xi_1}+\frac{f_{222}\xi_3}{\xi_2}-F_1\xi_2-F_2\xi_2+\frac{G_1}{2\xi_1}-\frac{G_2}{\xi_2} \right).
	\end{align}
\end{subequations}
The embedding space result \cite{Billo} also involves 12 structures.
\\

\noindent
If the rank-2 tensor $S_{\mu\nu}$ is actually the stress tensor $T_{\mu\nu}$ 
and the vector $V_\mu$ a conserved current
$J_\mu$, 
then we should furthermore set $\Delta=d$, $\Delta'=d-1$ and impose the conservation conditions.
%
%
Conservation of $J^\mu$ gives four PDE constraints while the conservation of $T^{\mu\nu}$ gives six constraints. 
(We simply count the structures needed to write down $\langle S_{\mu\nu} (x) \mathcal{O}(x')\rangle$
and $\langle V_\mu(x) V'_\nu(x') \rangle$, respectively and where $S$ is traceless.)
In total, there are 12 functions and 10 PDE relations, which we will spare the reader.
\paragraph{$\textbf{q}=\textbf{1}$:} Here $\Xi^{(2)}$, $\Xi'^{(2)}$, $\mathcal{J}$ and $\mathcal{J}'$ all vanish and the number of structure reduces to 2. The conservation of $T^{\mu \nu}$ gives two ODE constraints while the conservation of $J_{\mu}$ gives one ODE constraint. Since there are only 2 functions but 3 constraints, the system is overdetermined, and one might guess the correlation function vanishes. However, some of the constraints are degenerate, and the correlation function is fixed up to a constant  (see p 14 of \cite{Herzog1}). 
\paragraph{$\textbf{q}=\textbf{2}$:} Here $\mathcal{J}$ and $\mathcal{J}'$ are not independent and the number of structures reduces to 8. The conservation of $J^\mu$ now gives three PDE constraints while the conservation of $T^{\mu\nu}$ gives five PDE constraints. This is a system with 8 functions and 8 PDE relations.
It would be interesting to see if the system can be solved.

\subsection{$\langle TT \rangle$}
The two point correlator between symmetric rank-2 tensors $S$ and $S'$ is given by \eqref{symmetric2tensor}.
 There is a total of 36 independent components. Demanding that $S_{\mu \nu}$ is traceless gives 5 constraints and then demanding $S'_{\mu \nu}$ is traceless gives a further 4 constraints. This is a total of 9 constraints for 36 structures, hence reducing the number of independent structures to 27 and so matching the number predicted by the embedding formalism. For $\langle TT \rangle$ we need to further impose symmetry associated with identical operators alongside tracelessness. Symmetrising gives the two point correlation function of $T$, 
\begin{equation} \label{TTcorrelator}
\begin{split}
&\langle T_{\mu \nu}(x) T_{\alpha \beta}(x') \rangle = \frac{1}{s^{2d}}\bigg[\sum_{n=1}^{10}f_n(\xi_1,\xi_2)T^{(n)}_{\mu \nu; \alpha \beta}  + \sum_{m=1}^{9}g_m(\xi_1,\xi_2)S^{(m)}_{\mu \nu; \alpha \beta}  \\
&+h_1 \delta_{\mu \nu} \delta_{\alpha \beta} + h_2 \left(\delta_{\mu \nu}\Xi'^{(1)}_{\alpha}\Xi'^{(1)}_{\beta} + \delta_{\alpha \beta}\Xi^{(1)}_{\mu}\Xi^{(1)}_{\nu} \right) + h_3 \left(\delta_{\mu \nu}\Xi'^{(2)}_{\alpha}\Xi'^{(2)}_{\beta} + \delta_{\alpha \beta}\Xi^{(2)}_{\mu}\Xi^{(2)}_{\nu} \right) \\
&+h_4 \left( \delta_{\mu \nu}(\Xi'^{(1)}_{\alpha}\Xi'^{(2)}_{\beta} + \Xi'^{(1)}_{\beta}\Xi'^{(2)}_{\alpha}) + \delta_{\alpha \beta}(\Xi^{(1)}_{\mu}\Xi^{(2)}_{\nu} + \Xi^{(1)}_{\nu}\Xi^{(2)}_{\mu})  \right) \\
&+ H \left(\mathcal{J}_{\mu \nu}\delta_{\alpha \beta} + \delta_{\mu \nu}\mathcal{J}''_{\alpha \beta} \right) \bigg],
\end{split}
\end{equation}
where $T^{(n)}_{\mu \nu; \alpha \beta}, S^{(n)}_{\mu \nu; \alpha \beta}$ can be found in \eqref{q2structuresTT1} and \eqref{q2structuresTT2} and the functions $h_1, \ldots, h_4$ and $H$ satisfy the following relations,
\begin{subequations} \label{TTconstraints}
	\begin{align}
		h_1 &= -\frac{2f_1}{d}+\frac{4\xi_2}{d^2}\left(\frac{1}{u^2}+\frac{\xi_3}{2\xi_1} \right)f_2 -\frac{4\xi_3}{d^2}\left(1-\frac{\xi_3}{2\xi_1} \right)f_3 -\frac{4\xi_3}{d^2}\left(1+\frac{1}{u^2} \right)f_4 \\
	&+\frac{f_5}{u^4d^2} + \frac{\xi_3^2}{\xi_2^2d^2}f_6 +\frac{2\xi_3}{\xi_2u^2d^2}f_7- \frac{2\xi_3}{\xi_1u^2d^2}f_8 - \frac{2\xi_3^2}{\xi_1\xi_2d^2}f_9 + \frac{\xi_3^2f_{10}}{\xi_1^2d^2} + \frac{2}{d^2}\left((q-1)+\frac{\xi_3}{\xi_2} \right)g_1 \nonumber \\
	&-\frac{\xi_3}{\xi_1^2d^2}g_2-\frac{4\xi_3}{\xi_2^2d^2}g_3 + \frac{4\xi_3}{\xi_1\xi_2d^2}g_4 + \frac{2(q-1)}{u^2d^2}g_5 + \frac{2(q-1)\xi_3}{\xi_2d^2}g_6 - \frac{2(q-1)\xi_3}{\xi_1d^2}g_7 \nonumber \\ 
	&+ \frac{(q-1)^2}{d^2}g_8 +\frac{4}{d^2}\left((q-1)-\frac{\xi_3}{2\xi_1} \right)g_9, \nonumber \\
	h_2 &= -\frac{4\xi_2}{d}f_2+\frac{4\xi_3}{d}f_4 - \frac{f_5}{u^2d} -\frac{\xi_3}{\xi_2 d}f_7 + \frac{\xi_3}{\xi_1 d}f_8 - \frac{q-1}{d}g_5,\\
	h_3 &= \frac{4\xi_2}{d}f_3 + \frac{4\xi_2}{d}f_4 -\frac{\xi_3 }{\xi_2d}f_6 - \frac{f_7}{u^2d}+\frac{\xi_3}{\xi_1d}f_9-\frac{2g_1}{d}+\frac{4g_3}{\xi_2d}-\frac{2g_4}{\xi_1d}-\frac{q-1}{d}g_6,\\
	h_4 &= \frac{2\xi_2}{d}f_2 + \frac{2\xi_3}{d}f_3 - \frac{f_8}{u^2d} - \frac{\xi_3}{\xi_2d}f_9 + \frac{\xi_3}{\xi_1d}f_{10}-\frac{g_2}{\xi_1d}+\frac{2g_4}{\xi_2d} -\frac{q-1}{d}g_7-\frac{2g_9}{d},
	\end{align}
\end{subequations}
and,
\begin{equation}
H = -\frac{2}{d}g_1-\frac{g_5}{u^2 d}-\frac{\xi_3}{\xi_2 d}g_6+\frac{\xi_3}{\xi_1 d}g_7-\frac{q-1}{d}g_8-\frac{4}{d}g_9.
\end{equation}
 There are also additional constraints from conservation of $T$. This gives twelve PDE constraints and so we have a system with 19 functions and 12 PDE relations.
\paragraph{$\textbf{q}=\textbf{1}$:} In this case $\Xi^{(2)}=\Xi'^{(2)}=0$ and the independent structures reduce down to 5 with 2 constraints coming from tracelessness. This leaves a total of 3 independent structures appearing in 
$\langle T^{\mu\nu} (x) T^{\lambda\rho}(x') \rangle$. Conservation of $T^{\mu\nu}$ gives two ODE constraints and so we have a system of 3 functions with 2 ODE relations, reducing the number of independent functions to 1, as discussed long ago in \cite{McAvityOsborn2}. 
\paragraph{$\textbf{q}=\textbf{2}$:} In this case $\mathcal{J}$, $\mathcal{J}'$, and $\mathcal{J}''$ are not independent and so we set all the $g_m=0$ and $H=0$, leaving us with 14 structures. Furthermore, the traceless constraints also reduce down to four and hence $\langle T^{\mu\nu} (x) T^{\lambda\rho}(x') \rangle$ 
has 10 independent structures. These are given by the $T^{(n)}$ structures  in \eqref{q2structuresTT1} via \eqref{TTcorrelator} subject to \eqref{TTconstraints} with $g_m=H=0$.
(The $S^{(m)}$ structures vanish.)
Conservation of $T^{\mu\nu}$ gives eight PDE constraints, and so we have a system of 10 functions with 8 relations. 

\subsection{Constraints from Maxwell's Equation in $d=4$ on $\langle F \mathcal{O} \rangle$ and $\langle FF \rangle$}
\label{FF}
\subsubsection*{$\langle F \mathcal{O} \rangle$}
The correlation function between a Maxwell field strength and a scalar is given by,
\begin{equation}
\langle F_{\mu \nu}(x) \mathcal{O}(x') \rangle = \frac{|y'|^{2-\Delta'}}{(s^2)^2} f \left(\Xi^{(1)}_{\mu}\Xi^{(2)}_{\nu} - \Xi^{(1)}_{\nu}\Xi^{(2)}_{\mu}\right).
\end{equation}
Imposing the equation of motion $\partial_{\mu} F^{\mu\nu}=0$ and using the identity on \eqref{FOidentity}, we obtain two PDE constraints for $f$,
\begin{equation}
\begin{aligned}
\left(-3\xi_3 +2\xi_1 \left(\frac{\xi_3}{\xi_2} +q -1 \right)\right)f + \xi_1\xi_3 \left(f^{(1,0)}-2f^{(0,1)} \right) =0,
\end{aligned}
\end{equation}
and, 
\begin{equation}
\left(2\xi_1(2-q) +\xi_3 -2\xi_2\right)f + \frac{2\xi_1^2}{u^2}f^{(1,0)} -\xi_2\xi_3 f^{(0,1)} =0.
\end{equation}
The PDEs can be solved to find,\footnote{%
 Although we don't know what to make of it, it is interesting to note that 
 one can impose a massless free field equation on the scalar field as well, provided $q=3$ and $\Delta'=1$.
}
\begin{equation} \label{FOsolution}
f(\xi_1,\xi_2)= \frac{c \xi_1^3 \xi_2}{(1-\xi_2^2)^{\frac{q-1}{2}} \left(-1+(2\xi_1+\xi_2)^2 \right)^{\frac{p+1}{2}} },
\end{equation}
where $c$ is an integration constant, $p=d-q$ and the solution is valid for $p=1,q=3$ and $p=2,q=2$. 
The combination $-1+(2\xi_1+\xi_2)^2$ in the denominator of the expression diverges at the defect and goes to zero in the coincident limit.  
(For the case $q=1$,  $\Xi^{(2)}$ vanishes and the $\langle F \mathcal{O} \rangle$ correlator is automatically zero.)
\subsubsection*{$\langle FF \rangle$} 
The correlation function between two Maxwell field strength tensors is given by,
\begin{equation} \label{generalFF}
\begin{split}
\langle F_{\mu \nu}(x) &F_{\alpha \beta}(x') \rangle = \frac{4}{(s^2)^2} \bigg[\frac{f_1}{2} I_{\mu [\alpha}I_{\beta] \nu} + f_2 \Xi^{(1)}_{[\nu} I^{\phantom{(1)}}_{\mu] [\alpha} \Xi'^{(1)}_{\beta]} \\
+ &f_3\left( \Xi^{(1)}_{[\nu} I^{\phantom{(1)}}_{\mu] [\alpha} \Xi'^{(2)}_{\beta]} + \Xi^{(2)}_{[\nu} I^{\phantom{(1)}}_{\mu] [\alpha} \Xi'^{(1)}_{\beta]} \right) + f_4\Xi^{(2)}_{[\nu} I^{\phantom{(2)}}_{\mu] [\alpha} \Xi'^{(2)}_{\beta]} + f_5 \Xi^{(1)}_{[\mu}\Xi^{(2)}_{\nu]} \Xi'^{(1)}_{[\alpha} \Xi'^{(2)}_{\beta]} \\
+&\frac{f_6}{2} \mathcal{J}'_{\mu [\alpha|} \mathcal{J}'_{\nu |\beta]} + f_7 \Xi^{(1)}_{[\mu} \mathcal{J}'_{\nu] [\beta}\Xi'^{(1)}_{\alpha]} + f_8\left( \Xi^{(1)}_{[\mu} \mathcal{J}'_{\nu] [\beta} \Xi'^{(2)}_{\alpha]} + \Xi^{(2)}_{[\mu} \mathcal{J}'_{\nu] [\beta} \Xi'^{(1)}_{\alpha]} \right) \\
+ &f_9 \Xi^{(2)}_{[\mu} \mathcal{J}'_{\nu] [\beta}\Xi'^{(2)}_{\alpha]} + f_{10} \mathcal{J}'_{[\mu| [\alpha} I_{\beta] |\nu]} \bigg].
\end{split}
\end{equation}
Reflection positivity demands that,
\begin{equation} \label{reflectionpositivityonFF}
\begin{split}
2(f_1 + f_{10})(\xi_1,1) \geq 0, 
\quad
2(f_1 + 2f_{10} + f_6)(\xi_1,1) \geq 0,
\quad
2 \left(f_1 + \frac{f_2}{u^2}\right)(\xi_1,1) \geq 0, \\
\quad
2 \left(f_1 + f_{10} + \frac{f_2 + f_7}{u^2}\right)(\xi_1,1) \geq 0.
\end{split}
\end{equation}
\paragraph{$\textbf{q} = \textbf{3}$:} When $q=3$ we find that,
\begin{equation} \label{FFsimplificationq3}
\mathcal{J}'_{\mu [\alpha|} \mathcal{J}'_{\nu |\beta]} =- \frac{2}{\xi_3} \Xi^{(2)}_{[\mu} \mathcal{J}'_{\nu] [\beta} \Xi'^{(2)}_{\alpha]},
\end{equation}
and the number of independent structures reduces to nine. 
\paragraph{$\textbf{q} = \textbf{2}$:}
When $q=2$ we can set $f_6,...,f_{10}$ to zero due to \eqref{Jprimenotindp}. So, focusing on the $p=q=2$ case and applying the equation of motion gives four PDE constraints for $f_1,...,f_5$,

\begin{equation}
\begin{aligned}
&2\xi_1^2\xi_2 f_1^{(1,0)} -2\xi_1\xi_2\xi_3 f_1^{(0,1)} -2(\xi_1+2\xi_2)f_2 + \frac{2\xi_1^2}{u^2}f_2^{(1,0)} - \xi_2\xi_3 f_2^{(0,1)} \\
&+ \left(-\frac{2\xi_1}{\xi_2^2} +3\xi_3\right)f_3 -\xi_1\xi_3 \left(f_3^{(1,0)} -2 f_3^{(0,1)} \right) = 0,
\end{aligned}
\end{equation}

\begin{equation}
\begin{aligned}
&-2\xi_1^2 \xi_2 f_1^{(1,0)} -2\xi_1\xi_2^2 f_1^{(0,1)} +\xi_2 f_2 +(\xi_3 -\xi_2 -2\xi_1)f_3 + \frac{2\xi_1^2}{u^2}f_3^{(1,0)} -\xi_2\xi_3f_3^{(0,1)} \\
&-2\left(\frac{2\xi_1}{\xi_2^2}-\xi_3 \right)f_4 -\xi_1\xi_3 f_4^{(1,0)} +2\xi_1\xi_3 f_4^{(0,1)} = 0,
\end{aligned}
\end{equation}

\begin{equation}
\begin{aligned}
&-2\xi_1^2\xi_2 f_2^{(1,0)} -2\xi_1 \xi_2^2 f_2^{(0,1)} -\frac{2\xi_1}{\xi_2}(1+\xi_2^2)f_3 -2\xi_1^2\xi_2 f_3^{(1,0)} +2\xi_1\xi_2\xi_3 f_3^{(0,1)} + 2\xi_1\xi_3 f_4 \\
&-4\left(\frac{\xi_1}{\xi_2^2}-\xi_3\right)f_5 -\xi_1\xi_3 \left(f_5^{(1,0)}-2f_5^{(0,1)}\right) = 0,
\end{aligned}
\end{equation}

\begin{equation}
\begin{aligned}
&\frac{2\xi_1\xi_2^2}{\xi_3}f_2 + \frac{2\xi_1 \xi_2^2}{\xi_3}f_3 -2\xi_1^2\xi_2 f_3^{(1,0)} -2\xi_1\xi_2^2 f_3^{(0,1)} -\frac{4\xi_1}{\xi_2}f_4 -2\xi_1^2\xi_2f_4^{(1,0)} +2\xi_1\xi_2\xi_3f_4^{(0,1)} \\
&+2(\xi_2-\xi_3)f_5 -\frac{2\xi_1^2}{u^2}f_5^{(1,0)} + \xi_2\xi_3 f_5^{(0,1)} = 0.
\end{aligned}
\end{equation}
\paragraph{$\textbf{q} = \textbf{1}$:} When $q=1$, only $f_1$ and $f_2$ are present with the usual simplifications and there is only one ODE constraint,
\begin{equation}
2\xi_1^2\left(f_1'+\frac{f_2'}{v^2} \right)=4f_2.
\end{equation}
We may make this ODE look simpler by changing basis $f_2 \rightarrow v^2 g_2$ and by changing variables
 from $\xi_1 \rightarrow v$. This results in the ODE,
\begin{equation*}
v \frac{d}{dv}(f_1+g_2)=2g_2.
\end{equation*}
Since the coefficient of the derivative terms are equal, we may change basis again to $f_1 \rightarrow g_1- g_2$, which simplifies the ODE further,
\begin{equation}
g_2=\frac{v}{2} \frac{dg_1}{dv}.
\end{equation}
A free Maxwell theory in the bulk has only one independent structure for $q=1$. 


\section{Bulk-Defect Two Point Functions} \label{sec:bulktodefectlimit}
%
%
One reason to study bulk-defect two point functions is the fact that any bulk primary operator can be expressed
as a sum over operators on the defect -- the defect OPE.
The defect OPE of any bulk primary $\mathcal{O}_I(x)$ with dimension $\Delta_{\mathcal{O}}$ can be written as,
\begin{equation} \label{defectOPE}
\mathcal{O}_I(\textbf{x},y)= \sum_{\hat{\phi}_J} \sum_{\hat{\Delta}_{\hat{\phi}}}  \frac{1}{|y|^{\Delta_{\mathcal{O}} - \hat{\Delta}_{\hat{\phi}}}} \mathcal{D}_I^J(\vec{\partial},y,c^{(n)}_{\mathcal{O} \hat{\phi}}, \hat{c}_{\hat{\phi} \hat{\phi}}) \hat{\phi}_J(\textbf{x}),
\end{equation}
where $\hat{\phi}_J$ is a defect primary with dimension $\hat{\Delta}_{\hat{\phi}}$ and $\mathcal{D}_I^J$ is a parallel derivative operator depending on the bulk-to-defect coefficients $c^{(n)}_{\mathcal{O} \hat{\phi}}$ and the two point coefficient 
$\hat{c}_{\hat{\phi} \hat{\phi}}$
of $\hat{\phi}$. In particular, if the bulk-to-defect coefficients between $\mathcal{O}$ and some $\hat{\phi}$ are vanishing, then $\hat{\phi}$ cannot appear in defect OPE of $\mathcal{O}$.
%
To derive the contribution of a particular $\hat{\phi}_K$ to the defect OPE, we simply take the product with \eqref{defectOPE} and calculate the correlation function,
\begin{equation}
\frac{1}{|y|^{\Delta_{\mathcal{O}} - \hat{\Delta}_{\hat{\phi}}}} \mathcal{D}_I^J \langle \hat{\phi}_J(\textbf{x}) \hat{\phi}_K(\textbf{x}') \rangle = \langle \mathcal{O}_I(x) \hat{\phi}_K(\textbf{x}') \rangle = \frac{c^{(n)}_{\mathcal{O} \hat{\phi}}}{|y|^{\Delta_{\mathcal{O}} - \hat{\Delta}_{\hat{\phi}}}} \frac{\mathcal{T}_{(n)IK}}{(\textbf{s}^2 + |y|^2)^{\hat{\Delta}}},
\end{equation}
where $\mathcal{T}_{(n)}$ are the appropriate tensor structures. 
%
By matching the expansion on both sides as $y\rightarrow 0$, one can in principle determine the $\mathcal{D}_I^J$, although
our interest is not in these operators.\footnote{%
See \cite{McAvityOsborn1} for the $q=1$ case and \cite{Billo} for $q>1$.}
Instead, we focus on the precise form of bulk-defect two point functions and the tensor structures $\mathcal{T}_{(n)}$.

\subsection{Tensor Structures for the Defect Limit}
\begin{table}[h]
	\centering
	\begin{tabular}{|c|c|c|c|c|c|c|c|c|c|}
		\hline \hline \hline
		\multicolumn{10}{|c|}{\textbf{Independent Tensor Structures for Bulk-to-Defect}}                                                                                                                        \\ \hline \hline \hline
		\multicolumn{3}{|c|}{Vectors}                                   & \multicolumn{2}{c|}{\multirow{2}{*}{Bivector}} & \multicolumn{5}{c|}{2-Tensor}                                                        \\ \cline{1-3} \cline{6-10} 
		\multirow{2}{*}{at $x$} & \multicolumn{2}{c|}{at $\textbf{x}'$} & \multicolumn{2}{c|}{}                          & \multirow{2}{*}{at $x$}                      & \multicolumn{4}{c|}{at $\textbf{x}'$} \\ \cline{2-5} \cline{7-10} 
		& spin $=a$        & $i$          & $(\mu,a)$                   & $(\mu,i)$                 &                                              & (a,b)         &(a,i)   &(i,a)   &(i,j)         \\ \hline
		$\hat{\Xi}^{(1)}_{\mu}$                   & 0               & $\hat{\mathcal{X}}'_{i}$       & $\hat{\mathcal{I}}_{\mu a}$          & $\hat{\mathcal{I}}_{\mu i} $         & \multicolumn{1}{l|}{$\mathcal{J}_{\mu \nu}$, $\delta_{\mu \nu}$} & $\delta_{ab}$   & 0    & 0    & $\delta_{ij}$   \\ \hline \hline \hline
	\end{tabular}
	\caption{List of independent tensor structures used to construct bulk-to-defect two point correlation functions. The definition of these structures can be found in \eqref{bulktodefectstructures} where we use a hat on the bulk tensor structure to indicate it being evaluated at $y' = 0$. For a bulk-to-defect two point correlation function, the tensor structures at $x$ can have bulk Lorentz spin while the structures at $\textbf{x}'$ can either have parallel and/or orthogonal spin. Likewise, a bivector at $(x, \textbf{x}')$ will have one bulk spin and one parallel or orthogonal spin. Hence we separate the spin combinations in this table. }
	\label{bulktodefecttable}
\end{table}
\noindent
A defect of our formalism, which is not shared by the embedding space method,
is that in the absence of a cross ratio, we do not have a procedure for
constructing relevant tensor quantities.  We saw this issue in the case of the one-point function,
which must be constructed out of products of the rank-2 tensor 
${\mathcal J}_{\mu\nu}$.  We found the tensor ${\mathcal J}_{\mu\nu}$
in the process of constraining the two-point functions.  
The structure ${\mathcal J}_{\mu\nu}$ occurs in the contraction of ${\mathcal J}'_{\mu\alpha}$ with itself,
and the bivector ${\mathcal J}'_{\mu\alpha}$ in turn arises as a mixed derivative of the cross ratio $\xi_2$.  
Despite the fact that there is no cross ratio for the one-point function, 
having found ${\mathcal J}_{\mu\nu}$, we are free to use it in the construction of one-point functions.
\\

\noindent
The situation in the case of bulk-to-defect two point functions is similar.  
We have no cross ratio, but again there is a workaround.  
We can study the defect limit of bulk-bulk two point functions.  From this procedure,
we recover all of the relevant tensor structures necessary for constructing a 
general bulk-to-defect two-point function from scratch and the independent structures 
are given in table \ref{bulktodefecttable}.  We make this claim because there is 
a one-to-one match of the structures we find here to the structures
required in embedding space \cite{Billo}.
\\

\noindent
Here we analyse the limit where $y'=0$. 
An immediate issue is that the cross ratio $\xi_2$ and several of 
the tensor structures have an ill behaved $y' \to 0$ limit.
We remedy this problem by taking various linear combinations of 
the cross ratios and multiplying by appropriate powers of $\xi_1$ and $\xi_2$.
A basis with well defined limit as $y' \rightarrow 0$ is,
\begin{equation}\label{basisforboundarylimit}
\begin{split}
\left\{\Xi^{(1)}_{\mu}, \frac{\xi_2}{\xi_1}\Xi^{(2)}_{\mu} \right\} \& \left\{\mathcal{X}'_{\alpha}, \frac{\xi_2}{\xi_1^2}\Xi'^{(2)}_{\alpha} \right\}, 
\quad
\left\{\mathcal{I}_{\mu \alpha}, \bar{\mathcal J}'_{\mu \alpha}  \right\}, 
\quad
\left\{\delta_{\mu \nu}, {\mathcal J}_{\mu \nu} \right\} \& \left\{\delta_{\alpha \beta}, \bar{\mathcal J}''_{\alpha \beta}   \right\},
\end{split}
\end{equation}
for vectors at $x$ and $x'$, bivectors, and rank two tensors at $x$ and $x'$ respectively and where $\mathcal{X}'_{\alpha}:=\xi_2(\Xi'^{(1)}_{\alpha}-\Xi'^{(2)}_{\alpha})$, $\mathcal{I}_{\mu \alpha}:=I_{\mu\alpha} - \Xi^{(1)}_{\mu}\mathcal{X}'_{\alpha}$, $\bar{\mathcal{J}}'_{\mu \alpha} := \mathcal{J}'_{\mu \alpha} - \Xi^{(2)}_{\mu} \mathcal{X}'_{\alpha}$ and $\bar{\mathcal{J}''}_{\alpha \beta} := \mathcal{J}''_{\alpha \beta} + \Xi'^{(1)}_{\alpha} \Xi'^{(1)}_{\beta}$. 
In fact, several of these structures simply vanish or are not independent in the defect limit as $y' \rightarrow 0$,
\begin{equation}
\frac{\xi_2}{\xi_1}\Xi^{(2)}_{\mu}\to 0  \ , \; \;
\frac{\xi_2}{\xi_1^2}\Xi'^{(2)}_{\alpha} \to 0 \ , \; \;
\mathcal{X}'_{a} \to 0 \ , \; \;
\bar{\mathcal J}'_{\mu a} \to 0 \ , \; \;
\bar{\mathcal J}'_{\mu i} \rightarrow \mathcal{I}_{\mu i} \ , \; \;
\bar{\mathcal J}''_{\mathcal{A}} \to 0 \ , \; \;
\bar{\mathcal J}''_{ij} \to \delta_{ij} \
\end{equation}
where $\mathcal{A}$ represents the choices of spin $(a,b)$, $(a,i)$ and $(i,a)$.
As in the one-point function case, ${\mathcal J}_{\mu\nu}$ is not affected by the limit.
Not counting the Kronecker delta's, the remaining structures reduce to,
\begin{equation} \label{bulktodefectstructures}
\begin{aligned}
\Xi^{(1)}_{\mu} \big{|}_{y'=0} &= \begin{cases} 
\frac{2|y|}{\textbf{s}^2 + |y|^2} \textbf{s}_a & \mu=a, \\ 
\left(\frac{2|y|^2}{\textbf{s}^2 + |y|^2} -1 \right)n_i & \mu = i,
\end{cases} \\
\qquad
\mathcal{X}'_{i} \big{|}_{y'=0} &= -n_i,
\end{aligned}
\quad
\begin{aligned}
\mathcal{I}_{\mu b}\big{|}_{y'=0} &= \begin{cases} 
\delta_{ab} - \frac{2 \textbf{s}_a \textbf{s}_b}{\textbf{s}^2 + |y|^2}& \mu=a, \\ 
-\frac{2|y|  n_i \textbf{s}_b}{\textbf{s}^2 + |y|^2}& \mu = i,
\end{cases} \\
\qquad
\mathcal{I}_{\mu j}\big{|}_{y'=0}  &= \begin{cases} 
0& \mu=a, \\ 
\delta_{ij} - n_i n_j& \mu = i.
\end{cases}
\end{aligned}
\end{equation}
We represent the bulk-to-defect tensor structures (see table \ref{bulktodefecttable}) with a hat, 
e.g.~$\hat{\Xi}^{(1)}_{\mu} = \Xi^{(1)}_{\mu}|_{y'=0}$ etc. 
Similar to the bulk tensor structures (see \eqref{contractions1} and \eqref{contractions2}), the set of independent bulk-to-defect tensor structures are also closed under contraction. Leaving out the trivial contractions involving the $\delta$'s we have, 
\begin{equation} \label{defectcontractions}
\begin{aligned}
\hat{\Xi}^{(1)}_{\mu} \hat{\Xi}^{(1) \mu} &= \hat{\mathcal{X}}'_i \hat{\mathcal{X}}'^i =1, \\
\hat{\Xi}^{(1)}_{\mu} \hat{\mathcal{I}}^{\mu}_{\;\; a} &= \hat{\Xi}^{(1)}_{\mu} \hat{\mathcal{I}}^{\mu}_{\; \; i} = 0, \\
\hat{\mathcal{X}}'^i \hat{\mathcal{I}}^{\mu}_{\; \; i} &= \mathcal{J}_{\mu \nu} \hat{\mathcal{I}}^{\mu}_{\; \; a} = 0, \\
\hat{\Xi}^{(1)}_{\mu} \mathcal{J}^{\mu}_{\; \; \nu} &= \hat{\mathcal{I}}_{\mu i} \hat{\mathcal{I}}^{\mu}_{\; \; a} = 0, 
\end{aligned}
\qquad
\begin{aligned}
\mathcal{J}_{\mu \nu} \hat{\mathcal{I}}^{\mu}_{\; \; i} &= \hat{\mathcal{I}}_{\nu i}, \\
\hat{\mathcal{I}}_{\mu a} \hat{\mathcal{I}}^{\mu}_{\; \; b} &= \delta_{ab}, \\
\hat{\mathcal{I}}_{\mu i} \hat{\mathcal{I}}_{\nu}^{\; \; i} &= \mathcal{J}_{\mu \nu}, \\
\hat{\mathcal{I}}_{\mu i} \hat{\mathcal{I}}^{\mu}_{\; \; j} &= \delta_{ij} - \hat{\mathcal{X}}'_i \hat{\mathcal{X}}'_j 
= \hat{\mathcal I}_{ij}, \\
\hat{\mathcal{I}}_{\mu a} \hat{\mathcal{I}}_{\nu}^{\; \; a} &= \delta_{\mu \nu} - \hat{\Xi}^{(1)}_{\mu} \hat{\Xi}^{(1)}_{\nu} - \mathcal{J}_{\mu \nu}.
\end{aligned}
\end{equation}
We believe we have found all the relevant structures from which to construct the bulk-to-defect correlation function between operators of arbitrary spin.
Indeed, in embedding space, ref.\ \cite{Billo} construct these same bulk-to-defect two point functions from the five monomials
$Q_{BD}^i$, $i = 0, \ldots , 4$.  In our language, 
$Q^0_{BD}$ maps to ${\hat{\mathcal I}}_{\mu a}$, 
$Q^1_{BD}$ to ${\hat{\mathcal X}}'_i$, 
$Q^2_{BD}$ to $\hat{\Xi}^{(1)}_\mu$,
$Q^3_{BD}$ to ${\hat{\mathcal I}}_{\mu i}$,
and $Q^4_{BD}$ to ${\mathcal J}_{\mu\nu}$.
Given this mapping, we can see from a combinatorial point of view that exactly the same set of two-point functions should arise in both cases. In appendix \ref{bulktodefectlimitappendix}, we write a selection of bulk two point functions using the basis \eqref{basisforboundarylimit} in order to explore certain special cases where the bulk-to-defect coefficient can be obtained from the functions appearing in the bulk correlator. 

\subsection{$\langle V \hat{\mathcal{O}} \rangle$}
The two point correlation function between any bulk vector and any defect scalar is given by,
\begin{equation}
\label{VOdefect}
\langle V_\mu(x) \hat {\mathcal O}({\bf x}') \rangle = \frac{c_{V \hat {\mathcal{O}}}}{s^{2 \hat \Delta} |y|^{\Delta - \hat \Delta}} \hat \Xi^{(1)}_{\mu}
\end{equation}
%
where\footnote{We continue to use this convention throughout this section.} $\textbf{s}=\textbf{x}-\textbf{x}'$ and $s^2=\textbf{s}^2+|y|^2$. 
\paragraph{Conserved Current:} When $\partial_{\mu}V^{\mu}(x)=0$, then we get the constraints $\Delta=d-1$ and $\hat{\Delta} =d-q=p$. The first constraint is required for a bulk conserved current while the second implies if the dimension of the defect scalar $\hat{\Delta} \neq p$ then $c_{V \hat{\mathcal{O}}}=0$ when $V$ is conserved. In particular, the defect OPE of a bulk conserved current $J_{\mu}(x)$ can only have scalars of dimension $p$. This reduces to the bCFT result when $q=1$ and hence $p=d-1$.  
\paragraph{Bulk Limit:} We consider the special case where the defect OPE of a bulk operator $\mathcal{O}(\textbf{x}',y')$ expanded around $y \rightarrow 0$ has a finite leading order contribution from only one defect scalar primary $\hat{\mathcal{O}}$, i.e.~$\mathcal{O}(\textbf{x},0) = \hat{\mathcal{O}}(\textbf{x})$. For such a case, we can obtain $c_{V \hat{\mathcal{O}}}$ through the function $f_1$ present in the bulk-bulk correlator $\langle V_{\mu}(x) \mathcal{O}(x') \rangle$ in \eqref{VObulk1}. The relation is simply the boundary condition,
\begin{equation} \label{VObulklimit}
\begin{split}
\lim_{\xi_1 \rightarrow \infty}(4 \xi_1)^{\Delta'} f_1(\xi_1,\xi_2) =  c_{V\hat{\mathcal{O}}} , 
\qquad
\lim_{\xi_1 \rightarrow \infty}  \xi_1^{\Delta'+1} \xi_2^{-1} f_2(\xi_1,\xi_2) = \Fin . 
\end{split}
\end{equation}
Note the limit $\xi_1 \rightarrow \infty$ needs to exist independent of $\xi_2$ since $\xi_2$ has an undefined limit as $y' \rightarrow 0$.  
That the structure $\xi_2 \xi_1^{-1} \Xi_\mu^{(2)}$ vanishes in the defect limit places a finiteness constraint on $f_2$. 
\subsection{$\langle V \hat{V} \rangle$}
The two point correlation function between any bulk vector and any defect vector with parallel spin is given by,
\begin{equation}
 \label{VVparalleldefect}
\langle V_{\mu}(x) \hat{V}_a(\textbf{x}') \rangle =\frac{c_{V\hat{V}}}{s^{2 \hat \Delta} |y|^{\Delta-\hat{\Delta}} }
\hat {\mathcal I}_{\mu a} \ .
\end{equation}
%
%
\paragraph{Conserved Current:} When $\partial_{\mu}V^{\mu}(x)=0$, we simply get one condition $\Delta = d-1$ independent of $\hat{\Delta}$. If $\hat{V}$ is a conserved defect current and hence has $\hat \Delta = p-1$, the only way to satisfy the Ward identity $\partial_{a} \hat V^{a}(x)=0$ is to set $c_{V \hat V} = 0$.  In other words, a conserved defect current cannot appear in the defect OPE of any bulk vector.
\paragraph{Bulk Limit:} If there exists a $\hat{V}$ such that  $V'_a(\textbf{x}',0) = \hat{V}_a(\textbf{x}')$ we have the following boundary conditions on the functions appearing in \eqref{VVbulk}, 
\begin{equation} \label{VVparallelbulklimit}
\begin{split}
\lim_{\xi_1 \rightarrow \infty}g_s(\xi_1,\xi_2) = \Fin, 
\qquad
\lim_{\xi_1 \rightarrow \infty}g_5(\xi_1,\xi_2) = c_{V\hat{V}}, 
\end{split}
\end{equation}
where $s \in \{1,2,3,4,6\}$. To reiterate, finiteness of the basis \eqref{basisforboundarylimit} as $y' \rightarrow 0$ enables us to  put a finiteness condition on the bulk correlation functions directly, for this type of special case.
\\

\noindent
The two point correlation function between a bulk vector and a defect vector with orthogonal spin
can also be obtained from \eqref{VVbulk}:\footnote{%
	This result corrects a typo in (2.38) of \cite{Billo}.
}
\begin{equation}
 \label{VVorthogonaldefect}
\langle V_\mu(x) \hat{W}_j(\textbf{x}') \rangle = \frac{1}{s^{2\hat{\Delta}} |y|^{\Delta-\hat{\Delta}}}
\left( c_{V \hat W} \Xi^{(1)}_\mu \hat {\mathcal X}'_j + c'_{V\hat{W}} \hat {\mathcal I}_{\mu j}  \right)
\end{equation}
%
\paragraph{Conserved Current:} When $\partial_{\mu}V^{\mu}=0$, we get the constraints $\Delta=d-1$ and $c_{V\hat{W}}(\hat{\Delta}-p)=c'_{V\hat{W}}(q-1)$, reducing to one independent coefficient. In particular when $\hat{\Delta}=p$ and $q \neq 1$, we have $c'_{V\hat{W}}=0$.
%
%
When $q=1$, the $c'_{V\hat{W}}$ term vanishes since $\delta_{ij}-n_in_j \rightarrow 0$, and conservation states $c_{V\hat{W}}=0$ unless $\hat{\Delta}=p=d-1$; as expected, this is the same condition as for $\langle J \hat{\mathcal{O}} \rangle$.
%
%
Conversely, we see when $\hat{\Delta} =\Delta= d-1$  and $q \neq 1$, the Ward identity simply reduces to $c_{V \hat{W}} = c'_{V \hat{W}}$.
\paragraph{Bulk Limit:} If $V'_i(\textbf{x}',0) = \hat{W}_i(\textbf{x}') $, we have the following boundary conditions on the functions appearing in \eqref{VVbulk},
\begin{equation} \label{VVdefectrelation}
\begin{split}
\lim_{\xi_1 \rightarrow \infty}g_1(\xi_1,\xi_2) = c_{V\hat{W}}, 
\quad
\lim_{\xi_1 \rightarrow \infty}g_r(\xi_1,\xi_2) = \Fin, 
\quad
\lim_{\xi_1 \rightarrow \infty}(g_5 + g_6)(\xi_1,\xi_2) = c'_{V\hat{W}},
\end{split}
\end{equation} 
where $r\in \{2,3,4\}$. An example can be found in section \ref{freescalar}.
\subsection{$\langle S \hat{\mathcal{O}} \rangle$}
The correlation function between a bulk symmetric and traceless tensor with a defect scalar is given by, 
\begin{equation}
\langle S_{\mu\nu}(x) \hat{\mathcal{O}}(\textbf{x}')  \rangle = \frac{1}{s^{2\hat{\Delta}} |y|^{\Delta-\hat{\Delta}}} 
\left[
c_{S \hat O} \left( \hat \Xi^{(1)}_\mu \hat \Xi^{(1)}_\nu - \frac{1}{d} \delta_{\mu\nu} \right)
+ c'_{S \hat O} \left( {\mathcal J}_{\mu\nu} - \frac{q-1}{d} \delta_{\mu\nu} \right) 
\right]
\end{equation}
%
\paragraph{Energy Momentum Tensor:} When $S_{\mu \nu}=T_{\mu \nu}$ and hence $\partial_{\mu}T^{\mu}_{\nu}=0$ everywhere in the bulk, we obtain the usual constraint $\Delta=d$ and also the relation, 
\begin{equation}
\frac{\hat{\Delta}}{d}(q-1) c'_{S\hat{\mathcal{O}}}= - \left(\frac{\hat{\Delta}}{d}(1-d) +p \right)c_{S\hat{\mathcal{O}}},
\end{equation}
and so we only have one independent coefficient.  Similar to the $\langle V \hat V \rangle$ case we just considered,
when $q=1$, the structure multiplying $c'_{S\hat{\mathcal{O}}}$ is absent and $\hat{\Delta} = d$ (otherwise $c_{S \hat{\mathcal{O}}}=0$) meaning that the defect OPE of $T_{\mu \nu}$ can only have scalars with dimension $d$ (i.e.~the bCFT result). Likewise, when $\hat{\Delta}=d$ and $q \neq 1$ the Ward identity simply reduces to $c_{S \hat{\mathcal{O}}} = c'_{S \hat{\mathcal{O}}}$. 
\paragraph{Bulk Limit:} When $\mathcal{O}(\textbf{x}',0)=\hat{\mathcal{O}}(\textbf{x}')$, we have the following boundary conditions on the functions appearing in \eqref{SObulk},
\begin{equation} \label{SOdefectrelation}
\begin{split}
\lim_{\xi_1 \rightarrow \infty}g_1(\xi_1,\xi_2) = c_{S \hat{\mathcal{O}}}, 
\qquad
\lim_{\xi_1 \rightarrow \infty}g_n(\xi_1,\xi_2) = \Fin, 
\qquad
\lim_{\xi_1 \rightarrow \infty}g_4(\xi_1,\xi_2) = c'_{S\hat{\mathcal{O}}},
\end{split}
\end{equation}
where $n\in \{2,3\}$. We will see a counter example in section \ref{freescalar}.
\subsection{$\langle F \hat{V} \rangle$} \label{defectFV}
The correlation function between a bulk anti-symmetric tensor and a parallel spin defect vector is given by, 
\begin{equation}
\langle F_{\mu\nu}(x)\hat{V}_a(\textbf{x}') \rangle  = 
\frac{ 2c_{F\hat{V}}}{s^{2 \hat{\Delta}} |y|^{\Delta-\hat{\Delta}}}
\Xi^{(1)}_{[\mu} \hat {\mathcal I}^{\phantom{(1)}}_{\nu] a}
\end{equation}
%
We observe that for a point-like defect $p=1$ in four dimensions, 
where we additionally assume $\hat V_t = \hat J_t$ is a conserved charge,
we can interpret $\hat J_t$ as the insertion of a charge on the defect.  
Then the correlation function
\[
\langle F_{i t} J_t \rangle \sim \frac{ n_i }{|y|^2} 
\]
reduces to the statement of Coulomb's Law.  

\paragraph{Equation of Motion:} When $\partial_{\mu}F^{\mu}_{\nu}=0$ in the bulk (i.e.\ when the Maxwell field is free), we get the constraints $\Delta=d-2$ and $\hat{\Delta}=p-1$ and hence the defect vector has to be a conserved current meaning that the defect OPE of $F_{\mu \nu}$ which is free in the bulk can only contain defect vectors which are conserved. Furthermore, for a Maxwell theory we have $\Delta=d/2$ and we find that the equation of motion can only be applied when $d=4$. When $\hat{V}$ is a conserved current for any $F_{\mu\nu}$, we obtain the simple condition $\hat{\Delta}=p-1$. 
\paragraph{Bulk Limit:} When $V_a'(\textbf{x}',0) = \hat{V}_a(\textbf{x}') $ we have the following boundary conditions on the functions appearing in \eqref{FVbulk},
\begin{equation}
\begin{split}
\lim_{\xi_1 \rightarrow \infty}g_m(\xi_1,\xi_2) = \Fin, 
\qquad
\lim_{\xi_1 \rightarrow \infty}g_3(\xi_1,\xi_2) = c_{F \hat{V}},
\end{split}
\end{equation}
where $m\in \{1,2,4,5,6\}$.
\subsection{$\langle F \hat{F} \rangle$}
The correlation function between a bulk anti-symmetric tensor and a defect anti-symmetric orthogonal tensor is given by,
\begin{eqnarray}
\langle F_{\mu\nu}(x) \hat{F}_{ij}(\textbf{x}') \rangle &=& 
\frac{2}{s^{2 \hat \Delta} |y|^{\Delta - \hat \Delta}}
\Bigl[
c_{F \hat F} (\hat {\mathcal I}_{\mu i} \hat {\mathcal I}_{\nu j} -\hat {\mathcal I}_{\mu j} \hat {\mathcal I}_{\nu i} )
\nonumber \\
&&
\qquad \qquad + c'_{F \hat F} \left(  \hat {\mathcal X}'_i \hat \Xi^{(1)}_{[\mu} \hat {\mathcal I}^{\phantom{(1)}}_{\nu] j} - 
 \hat {\mathcal X}'_j \hat \Xi^{(1)}_{[\mu} \hat {\mathcal I}^{\phantom{(1)}}_{\nu] i} \right)
\Bigr]
\end{eqnarray}
%
%
%
\paragraph{Equation of Motion:} When $\partial_{\mu}F^{\mu}_{\nu}=0$ in the bulk, we obtain the condition $\Delta = d-2$ and the relation,
\begin{equation} \label{FFdefectWard}
c'_{F\hat{F}}(\hat{\Delta}-p)=-2(2-q)c_{F\hat{F}},
\end{equation}
leaving only one independent coefficient. When $q=2$ and $\hat{\Delta} \neq p$ we see that $c'_{F\hat{F}}=0$ and also the structure involving $c_{F\hat{F}}$ vanishes. Hence the correlation function is zero. If instead $\hat{\Delta}=p$ and $q \neq 2$, we see that $c_{F\hat{F}}=0$. In the special case when both $q=2$ and $\hat{\Delta}=p=d-2=\Delta$ the relation is automatically satisfied and we seem to have two independent coefficients. However, since for $q=2$ the structures corresponding to $c_{F\hat{F}}$ vanishes, we only have $c'_{F\hat{F}}$. Finally, when $\hat{\Delta}=\Delta=d-2$, irrespective of $q$ and $p$, the Ward identity simply reduces to $c'_{F \hat{F}} = 2c_{F \hat{F}}$. 
\paragraph{Bulk Limit:} When $ F'_{ij}(\textbf{x}',0) = \hat{F}_{ij}(\textbf{x}')$, we have the following boundary conditions on the functions appearing in \eqref{FFbulk},
\begin{equation} \label{FFdefectrelations}
\begin{split}
\lim_{\xi_1 \rightarrow \infty}g_m = \Fin, 
\qquad
\lim_{\xi_1 \rightarrow \infty} \left(g_1+\frac{g_7}{2}+g_{12} \right) = c_{F \hat{F}},
\qquad
\lim_{\xi_1 \rightarrow \infty}(g_2+g_8) = c'_{F \hat{F}},
\end{split}
\end{equation}
where $m \in \{3,4,5,6,9,10,11\}$. An example can be found in section \ref{FFdefectexample}.
\\

\noindent
Here are some more correlators which we list without further analysis, 
\begin{subequations}
	\begin{align}
	\langle \mathcal{O}(\textbf{x},y) \hat{W}_j(\textbf{x}') \rangle &= - \frac{c_{\mathcal{O}\hat{W}}}{(s^2)^{\hat{\Delta}} |y|^{\Delta-\hat{\Delta}}} n_j, \label{scalar-transversevector}\\
	\langle \mathcal{O}(\textbf{x},y) \hat{S}_{ij}(\textbf{x}') \rangle &= \frac{c_{\mathcal{O}\hat{S}}}{(s^2)^{\hat{\Delta}} |y|^{\Delta-\hat{\Delta}}} \left(n_i n_j - \frac{\delta_{ij}}{q} \right),
	\end{align}
\end{subequations}
where $\hat{S}_{ij}$ is a traceless and symmetric defect tensor.
\section{Free Field Defect CFT} \label{sec:free}
We will now focus on specific examples of free theories symmetric under the restricted conformal group \eqref{restrictedconf}. The purpose of looking at a free theory is mostly to study a simple example which has a defect symmetry allowing us to verify the general results in sections \ref{bulktwopointcorrelation} and \ref{sec:bulktodefectlimit}.

\subsection{Free Scalar Theory on $\mathbb{R}^p \times ({\mathbb R}^q / {\mathbb Z}_2)$} \label{freescalar}
The parent theory is $\mathcal{N}$ massless scalar fields in $d$ dimensions with the free propagator $\frac{\delta_{AB}}{(s^2)^{\Delta}}$, where $\Delta=\frac{d}{2}-1$. 
We then take an orbifold, where we identify directions normal to the defect $y \sim -y$.  The spacetime becomes 
$\mathbb{R}^p \times  ({\mathbb R}^q / {\mathbb Z}_2)$, and we are faced with a choice what to do to $\phi_A$ under the orbifold action.  Two natural choices are to send $\phi_A \to \pm \phi_A$.  The method of images then produces the propagator
\begin{equation} \label{scalarprop}
\langle \phi_A(x) \phi_B(x') \rangle = \delta_{AB} \left(\frac{1}{(s^2)^{\Delta}} + \frac{\lambda}{(\tilde{s}^2)^{\Delta}} \right) ,
\end{equation} 
where $\tilde{s}^2=(\textbf{x}-\textbf{x}')^2+(y+y')^2$ and $\lambda = \pm 1$.  In the codimension one case, these choices correspond to more familiar Neumann and Dirichlet boundary conditions.  In fact, letting $\lambda \in \mathbb{R}$  be arbitrary, the correlators have the appropriate symmetry for a dCFT with a $p$-dimensional defect.\footnote{ At least perturbatively, one can access more general values of $\lambda$ by including degrees of freedom on the boundary that interact with the scalar \cite{Herzog1}.
}  However, we will see shortly that $\lambda > 0$ violates an energy condition.
\\

\noindent
This theory exhibits a global $O(\mathcal{N})$ symmetry giving rise to the bulk conserved current,
\begin{equation}
J^{AB}_{\mu}=\phi^A\partial_{\mu}\phi^B-\phi^B\partial_{\mu}\phi^A,
\end{equation}
and from translation invariance (in the bulk) we have an improved energy-momentum tensor,
\begin{equation}
T_{\mu \nu}=\partial_{\mu}\phi_A \partial_{\nu}\phi^A - \frac{1}{2}\delta_{\mu \nu}\partial^{\alpha}\phi_A \partial_{\alpha} \phi^A - \frac{d-2}{4(d-1)}(\partial_{\mu}\partial_{\nu}-\delta_{\mu\nu} \partial_{\alpha}\partial^{\alpha})\phi^2,
\end{equation}
which has a non-zero one point function respecting the conformal symmetry given by, 
\begin{equation}
\langle T_{\mu \nu} (x) \rangle = -\frac{\lambda \mathcal{N}}{(2y)^d} \frac{d(d-2)}{(d-1)} \left[\mathcal{J}_{\mu \nu} - \frac{q-1}{d}\delta_{\mu \nu}\right].
\end{equation}
Curiously, 
positive $\lambda$ is not consistent with the Average Null Energy Condition (ANEC) when $q>1$.  
The ANEC states that the integral of the tangential-tangential component of the stress-tensor along a light-like trajectory must be positive.  
Appropriately Wick rotating our result to Lorentzian signature, 
that integral in this case is proportional to $\lambda$  \cite{Jensen:2018rxu}.  Of course in the codimension one case, 
$\langle T_{\mu\nu} \rangle$ itself vanishes and there is no such constraint.
The ANEC was proven to hold assuming a ${\mathbb R}^{d-1,1}$ space-time
\cite{Faulkner:2016mzt,Hartman:2016lgu}, and it is not clear that the theorem should hold in this more general orbifolded context.
\\

\noindent
Using \eqref{scalarprop} we find that the two point correlator between the conserved current respects the defect conformal symmetry,
\begin{equation} \label{JJproject}
\begin{split}
\langle J^{AB}_{\mu}(x) J^{CD}_{\nu}(x') \rangle= \frac{2}{(s^2)^{d-1}} \bigg(&f_1\Xi^{(1)}_{\mu} \Xi'^{(1)}_{\nu} 
+ f_2\Xi^{(2)}_{\mu} \Xi'^{(2)}_{\nu} + f_3(\Xi^{(1)}_{\mu} \Xi'^{(2)}_{\nu}+ \Xi^{(2)}_{\mu} \Xi'^{(1)}_{\nu})\\
\qquad
&+f_4 I_{\mu \nu} + f_5 {\mathcal J}'_{\mu \nu} \bigg) (\delta^{AC} \delta^{BD} - \delta^{AD} \delta^{BC}),
\end{split}
\end{equation}
where, 
\begin{equation}\label{JJprojective}
\begin{aligned}
f_1 =f_2 =-f_3 &=-2\lambda(d-2)\xi_2(u^2)^{\frac{d}{2}}\left(1+ \lambda (u^2)^{\frac{d}{2}}+\frac{d}{2}(u^2-1) \right), \\
\qquad
f_4 &= (d-2)\left(1+ \lambda (u^2)^{\frac{d}{2}-1} \right)\left(1+ \lambda(u^2)^{\frac{d}{2}}\right), \\
\qquad
f_5 &=-2\lambda(d-2)\left(1+ \lambda(u^2)^{\frac{d}{2}-1}\right)(u^2)^{\frac{d}{2}}.
\end{aligned}
\end{equation}
It can be shown that these functions satisfy the conservation PDEs \eqref{JJPDE1} and \eqref{JJPDE2}.

\subsubsection*{Defect Limit: $y' \rightarrow 0$}
If we change the basis to the one corresponding to \eqref{VVbulk}, the functions $g_i$ are given by,
\begin{equation}
\begin{aligned}
g_1 &= (d-2) \left(1- \lambda (u^2)^{\frac{d}{2}-1} (u^2 -1)(1+d u^2) - \lambda^2 (u^2)^{d-1}(-1 +2 u^2)\right),\\
g_4 &= - \lambda (d-2)  (u^2)^{\frac{d}{2}} \left(d u^2 + 2 \lambda (u^2)^{\frac{d}{2}} \right), \\
g_5 &= (d-2)\left(1+ \lambda (u^2)^{\frac{d}{2}-1} \right)\left(1+ \lambda(u^2)^{\frac{d}{2}}\right), \\
g_6 &= -2\lambda(d-2)\left(1+ \lambda(u^2)^{\frac{d}{2}-1}\right)(u^2)^{\frac{d}{2}}, \\
g_2 &= g_3 = 0. 
\end{aligned}
\end{equation}
We see that the $g_i$'s are all finite as $\xi_1 \rightarrow 0$. This implies the existence of a defect primary $\hat{W}_i(\textbf{x})$ such that the bulk-to-defect coefficients for $\langle J_{\mu}\hat{W}_i \rangle$ are given by \eqref{VVdefectrelation},
\begin{equation}
c_{V\hat{W}} = c'_{V\hat{W}} = (d-2)(1-\lambda^2).
\end{equation}
The Ward identity is also automatically satisfied by setting $\hat{\Delta} =d-1$. So, we see that for this free theory $J^{AB}_i(\textbf{x},0)$ is a defect primary. Furthermore, reflection positivity \eqref{boundonJJ} demands that $c'_{V\hat{W}} \geq0$ and hence we have a bound on $\lambda$, that $|\lambda| \leq 1$. 
\\

\noindent
For this theory, $\phi^2=\phi_A\phi^A$ is a bulk scalar primary. The one point and two point functions of $\phi^2$ obey the defect conformal symmetry and are given by,
\begin{equation}
\langle \phi^2(x) \rangle = \frac{\lambda \mathcal{N}}{(2|y|)^{d-2}},
\end{equation}
and
\begin{equation}
\langle \phi^2(x)\phi^2(x') \rangle= \frac{1}{(s^2)^{d-2}} \left[\lambda^2 \mathcal{N}^2 \xi_1^{d-2} + 2\mathcal{N} (1+ \lambda u^{d-2})^2 \right].
\end{equation}
From this we can read off the scaling dimension of $\phi^2$ to be $\Delta_{\phi^2}=d-2$. Using that $\phi^2$ is a primary, we can then look at its two point correlation function with the energy-momentum tensor. 
This two point function is consistent with conformal symmetry and is given by, 
\begin{equation} \label{TTproject}
\begin{split}
\langle T_{\mu \nu}(x) \phi^2(x') \rangle = -\frac{|y'|^{2 }}{s^{2d}} &\bigg[g_1\left(\Xi^{(1)}_{\mu}\Xi^{(1)}_{\nu}-\frac{\delta_{\mu \nu}}{u^2 d}\right) + g_2\left(\Xi^{(2)}_{\mu}\Xi^{(2)}_{\nu} - \frac{\delta_{\mu \nu} \xi_3}{\xi_2 d}\right) \\
+&g_3\left(2\Xi^{(1)}_{(\mu}\Xi^{(2)}_{\nu)} + \frac{\delta_{\mu \nu} \xi_3}{\xi_1 d} \right) + g_4\left(\mathcal{J}_{\mu\nu}-\frac{q-1}{d}\delta_{\mu \nu} \right) \bigg],
\end{split}
\end{equation}
where,
\begin{equation} \label{TOprojective}
g_1 =g_2 = -g_3 = 4\lambda \mathcal{N} \frac{d(d-2)^2}{(d-1)} (u^2)^{\frac{d}{2}+1}\xi_2^2 \ , \\
\qquad
g_4 = 4 \lambda^2\mathcal{N}^2 \frac{d(d-2)}{(d-1)} \xi_1^d.  \\
\end{equation}
It can be checked that these functions satisfy the conservation PDEs (\ref{firstTO}) and (\ref{secondTO}) 
for $\langle T \mathcal{O} \rangle$. This correlator provides a counter example to the relations on \eqref{SOdefectrelation} since $g_4$ diverges. This is due to the presence of the identity in the defect OPE of $\phi^2$ whose contribution to the OPE is singular as $y \rightarrow 0$.
Note that the two point correlator with a single $J^{AB}$ will always be zero since the $AB$ indices are  antisymmetric.
\subsection{Free Maxwell Theory on a Wedge}
Next, we consider a free U(1) gauge theory in four dimensions with a two dimensional orbifold defect.  The orbifold in this case is obtained by identifying the transverse ${\mathbb R}^2$ under rotations by $2 \pi / N$ for $N \geq 2$ an integer.  
We parametrise the space ${\mathbb R}^2 \times ({\mathbb R}^2 / {\mathbb Z}_N)$ by $x=(\textbf{x},y_1,y_2)$.
In the unorbifolded space, the free propagator in Feynman gauge is
\begin{equation}
\langle A_{\mu}(x) A_{\nu}(x') \rangle = \frac{\delta_{\mu \nu}}{s^2}.
\end{equation}

\noindent
We obtain the propagator on the orbifolded space by using the method of images.  
The sum is complicated to evaluate for general values of $N$ for arbitrary points $x$ and $x'$. 
General results for the sum looking at special points exist for the free scalar and spinor field theory on a wedge \cite{Herzog2,Herzog:2015cxa}, but not to our knowledge for the Maxwell field.
We will content ourselves by showing this Maxwell theory on a wedge is a (4,2) defect CFT in just two examples, $N=2$ and $N=4$.
\\

\noindent
Unlike the scalar case, the Maxwell field has a space-time index on which the rotation acts in a nontrivial fashion.  Like the scalar case, we have to decide further if the rotation gives an extra $\pm 1$ phase when acting on the Maxwell field (the analog of absolute and relative boundary conditions in the $q=1$ case).   
The action of the rotation on the space-time index is straightforward to work out in polar coordinates, $\hat{x}=(\textbf{x},r,\theta)$.
The extra phase we incorporate by introducing real parameters $\lambda$ and $\lambda'$.  In fact, we will see that the correlator has the correct dCFT form for general $\lambda$ and $\lambda'$ and not just for the special values $\pm 1$.

\subsubsection*{Wedge with $N=2$}
When $N=2$, 
we are working with a special case, 
$\mathbb{R}^2 \times ({\mathbb R}^2 / {\mathbb Z}_2)$,
 of the orbifold used for the scalar theory above.  
The propagator on the wedge is given by,
\begin{equation}
 \label{Nequals2prop}
G^{\lambda}_{\mu \nu}(x,x') = \frac{\delta_{\mu\nu}}{s^2} + \lambda \frac{M_{\mu \nu}}{\tilde{s}^2},
\end{equation}
where $M_{\mu \nu} = \Diag\{1,1,-1,-1\}$. The correlation function $\langle FF \rangle$ is now specified by the functions,
\begin{equation} 
 \label{wedge2}
\begin{aligned}
f_1 &=4(1+ \lambda u^4), \\
\qquad
f_2 &=-8 \lambda u^6\xi_2, \\
\qquad
f_3 &=8\lambda u^6\xi_2, 
\end{aligned}
\qquad
\begin{aligned}
f_4 &= -8\lambda u^6\xi_2+\frac{8\lambda u^4}{\xi_3}, \\
\qquad
f_5 &= - \frac{16\lambda u^6\xi_2}{\xi_3},
\end{aligned}
\end{equation}
on \eqref{generalFF}. The free Maxwell theory has a bulk stress tensor, $T_{\mu \nu}(x)=\tensor{F}{_\mu ^\gamma}\tensor{F}{_\nu_\gamma}-\frac{\delta_{\mu\nu}}{4} F_{\alpha \beta}F^{\alpha \beta}$, which has a non-zero one point function,
\begin{equation}
\langle T_{\mu \nu}(x) \rangle = -\frac{\lambda}{y^4} \left(\mathcal{J}_{\mu \nu} - \frac{1}{4}\delta_{\mu\nu}\right).
\end{equation}
As for the scalar, this result is consistent with the ANEC only for $\lambda \leq 0$. 
\subsubsection*{Wedge with $N=4$}
The propagator in this case on $\mathbb{R}^2 \times  ({\mathbb R}^2 / {\mathbb Z}_4)$ is given by the matrix,
\begin{equation}
G^4_{\mu \nu} = \begin{pmatrix}
\frac{1}{s^2}+\frac{\lambda}{\tilde{s}^2}+\frac{\lambda'}{s_+^2}+\frac{\lambda'}{s_{-}^2} & 0 & 0 & 0 \\
0 & \frac{1}{s^2}+\frac{\lambda}{\tilde{s}^2}+\frac{\lambda'}{s_+^2}+\frac{\lambda'}{s_{-}^2} & 0 & 0 \\
0 & 0 & \frac{1}{s^2}-\frac{\lambda}{\tilde{s}^2} & \frac{\lambda'}{s_+^2}-\frac{\lambda'}{s_{-}^2} \\
0 & 0 &  -\frac{\lambda'}{s_+^2}+\frac{\lambda'}{s_{-}^2} &   \frac{1}{s^2}-\frac{\lambda}{\tilde{s}^2} 
\end{pmatrix},
\end{equation}
%
where $s_+^2=\textbf{s}^2 + (y_1-y_2')^2 +(y_2+y_1')^2$ and $s_-^2=\textbf{s}^2 + (y_1+y_2')^2 +(y_2-y_1')^2$. We also define new functions of the cross ratios, $\xi^{\pm}_3=\frac{1}{2}(\xi_2 \pm \sqrt{1-\xi_2^2})$ and $u_{\pm}^2=\frac{\xi_1}{\xi_1 + \xi^{\pm}_3}$. Calculating $\langle FF \rangle$ we find that it again matches the general result on \eqref{generalFF} with the functions $f_1,...,f_5$ given by,
\begin{equation} \label{wedge4}
\begin{aligned}
f_1 &=4(1+\lambda u^4+\lambda' u_+^4+ \lambda' u_-^4), \\
\qquad
f_2 &=-8(\lambda u^6 \xi_2+ \lambda' u_+^6 \xi^+_3+ \lambda' u_-^6 \xi^-_3), \\
\qquad
f_3 &= -f_2+\lambda' g(u_+^2+u_-^2)^3, 
\qquad
\end{aligned} 
\begin{aligned}
f_4 &=-f_3 + \frac{8\lambda u^4}{\xi_3}+\lambda' h(u_+^2+u_-^2)^3, \\
\qquad
f_5 &=\frac{2f_2}{\xi_3}-2\lambda' g\xi_2(u_+^2+u_-^2)^3.
\end{aligned}
\end{equation}
where,
\begin{subequations}
	\begin{align}
	g &=\frac{2 \left(6 \xi_1^2+6 \xi_1 \xi_2+\xi_2^2\right)+1}{(2 \xi_1+\xi_2)^3}, \\
	h &=-\frac{\xi_2 (\xi_1+\xi_2) \left(4 \xi_1^2+4
		\xi_1\xi_2-2 \xi_2^2+3\right)}{\xi_1
		(\xi_2^2-1)(2 \xi_1+\xi_2)^2}.
	\end{align}
\end{subequations}

\noindent
The stress tensor one point function takes the usual form
\begin{equation}
\langle T_{\mu \nu}(x) \rangle = -\frac{\lambda + 8 \lambda'}{y^4} \left(\mathcal{J}_{\mu \nu} - \frac{1}{4}\delta_{\mu\nu}\right).
\end{equation}
Consistency with the ANEC would require 
the weaker constraint $\lambda + 8 \lambda' \leq 0$, which eliminates the particular choice 
$\lambda = -1$ and $\lambda' = 1$.  
\\

\noindent
Although we don't explicitly calculate $\langle FF \rangle$ for more general $N$, we expect this wedge theory to obey the conformal constraints for any positive integer $N$. The particular functions $f_1,...,f_5$ in \eqref{wedge2} and \eqref{wedge4}  for the wedge case above can also be shown to satisfy the four PDE constraints in section \ref{FF}.
\subsubsection*{$\langle T_{\mu \nu}(x) T_{\rho \sigma}(x') \rangle$}
 Given (for any $N$) a free Maxwell theory on a wedge that is conformal, the connected correlation function between two energy momentum tensors can be written using just the functions $f_1,...,f_5$ appearing in $\langle FF \rangle$. The result is \eqref{TTcorrelatorappendix}. 

\subsection{Maxwell Theory on a $\mathbb{R} \times  ({\mathbb R}^3 / {\mathbb Z}_2)$}  \label{FFdefectexample}
Another way to generalise the propagator \eqref{Nequals2prop} is to set $p=1$ and $q=3$. In doing so, we arrive at 
\begin{equation}
\bar{G}^2_{\mu \nu}=\frac{\delta_{\mu \nu}}{s^2} + \lambda \frac{\bar{M}_{\mu\nu}}{\tilde{s}^2},
\end{equation}
where $\bar{M}_{\mu \nu} = \Diag\{1,-1,-1,-1\}$ and our spacetime points are $x=(x_1,y_1,y_2,y_3)$. 
Calculating the correlator $\langle FF \rangle$, we find that it respects conformal symmetry and is given by \eqref{generalFF}, where the functions are,
\begin{equation} \label{FFNequals2withlambdaq3}
\begin{aligned}
f_1 &=4(1+ \lambda u^4), \\
\qquad
f_2 = f_4 = -f_3 &=-8 \lambda u^6\xi_2, \\
\qquad
f_5 = f_9 &= 0,
\end{aligned}
\qquad
\begin{aligned}
f_6 &= 16 \lambda u^4(1-u^2 \xi_2 \xi_3), \\
\qquad
f_7 = -f_8 &= 16 \lambda u^6 \xi_2, \\
\qquad 
f_{10} &= -8\lambda u^4.
\end{aligned}
\end{equation}
Note that since $q=3$, we used \eqref{FFsimplificationq3} to remove $f_9$ leaving us with only  nine independent structures instead of ten. In this case $\langle T_{\mu \nu}(x) \rangle =0$ and so there is no constraint on $\lambda$ from the ANEC.

\subsubsection*{Defect Limit: $y' \rightarrow 0$}
If we change the basis to the one corresponding to \eqref{FFbulk}, the functions $g_i$ are given by, 
\begin{equation}
\begin{aligned}
g_1 & = 2(1+\lambda u^4), \\
\qquad
g_2 &= 4[1+\lambda u^4(1-2u^2)], \\
\qquad
g_4 &= -8 \lambda u^6, \\
g_3 = g_5 = g_6 &= 0,
\end{aligned}
\qquad
\begin{aligned}
g_7 &= 16 \lambda u^4, \\
\qquad
g_8 &= 8\lambda u^4(2u^2-1), \\
\qquad 
g_{10} &= 16 \lambda u^6, \\
g_{12} &= - 8 \lambda u^4, \\
g_9 &= g_{11} = 0.
\end{aligned}
\end{equation}
We see that the $g_i$'s are finite as $\xi_1 \rightarrow \infty$. This implies the existence of a defect primary $\hat{F}_{ij}(\textbf{x})$ such that the bulk-to-defect coefficients for $\langle F_{\mu\nu} \hat{F}_{ij} \rangle$ are given by \eqref{FFdefectrelations},
\begin{equation} \label{FFdefectforwedge}
\begin{split}
c_{F\hat{F}}= 2(1+\lambda), 
\qquad 
c'_{F\hat{F}} = 4(1+\lambda).
\end{split}
\end{equation}
The Ward identity \eqref{FFdefectWard} is also satisfied as it reduces to $c'_{F\hat{F}} = 2 c_{F\hat{F}}$ when $\Delta' =2$, $p=1$ and $q=3$ . So, for this theory $F_{ij}(\textbf{x},0)$ is a defect primary. Reflection positivity demands that both $c_{F\hat{F}}$ and $c'_{F\hat{F}}$ are greater than zero, placing a lower bound $\lambda \geq -1$.

\section{Conclusion and Further Discussion}
In this paper we provide the necessary tensor structures, in configuration space, required to construct bulk two-point correlation functions for a conformal field theory with a flat defect.  In doing so, we 
find the appropriate tensor structures required for constructing any one-point and bulk-to-defect two point function as well.  
We further examine the conservation constraints on correlators involving a conserved current and the stress tensor.  We also looked at the free field constraints on correlators involving a Maxwell field.   These constraints are summarized in figure \ref{bulkcorrelatorlist}.  
\\

\noindent
While we did not provide detailed constructions, we believe, based on the discussion in section \ref{sec:higherpoint}, that it is straightforward to extend our result to higher point correlation functions.  One simply duplicates the tensor structures in figure \ref{Table:tensorstructures} for each pair of points and uses these as building blocks for the multi-point functions.  It would be interesting to explore these higher point cases further, and as a starting point to check that 
the relevant structures form a complete set, which we did in the two-point case but not in general.
\\

\noindent
Given the generally large number of undetermined functions required to specify our two-point functions, 
the particularly harsh constraints on a couple of our correlation functions call out for further analysis.  In analyzing 
$\langle F_{\mu\nu}(x) {\mathcal O}(x') \rangle$, we found that applying a free field constraint to $F_{\mu\nu}$ meant that the correlation function was determined up to a constant for theories with $d=4$ and $q=2$ or 3.  We would like to find an example where such a correlation function can be calculated and the result is nontrivial, i.e.\ not zero.
We also found that in codimension $q=2$ and for $\langle T_{\mu\nu}(x) J_\lambda(x') \rangle$ where $T_{\mu\nu}(x)$ is the stress tensor and $J_\lambda(x')$ a conserved current, the correlation function is fixed up to eight functions of two cross ratios that furthermore satisfied eight partial differential equations.   In other words, if the correlation function is specified along a particular slice in cross ratio space, it should generically be defined everywhere through the conservation equations.
\\

\noindent
A couple of other results are worthy of remark.  We analyzed the constraints of reflection positivity on the
$\langle J_\mu(x) J_\nu(x') \rangle$ and $\langle F_{\mu\nu} (x) F_{\lambda \rho}(x') \rangle$ correlation functions, 
finding (\ref{boundonJJ}) and (\ref{reflectionpositivityonFF}).  
(The structure of the $\langle T_{\mu\nu}(x) T_{\lambda \rho}(x')\rangle$ correlator is complicated enough that we
leave an analysis of reflection postivity of this structure for the future.)
We also found that some of our orbifold theories failed to
satisfy the ANEC.  As the ANEC was proven only for Lorentz invariant theories \cite{Faulkner:2016mzt,Hartman:2016lgu}, 
the result is intriguing but not in violation of
the theorem.
\\

\noindent
We have a particular interest in defect theories that are free in the bulk and have interactions confined to the defect.
In this set, perhaps the simplest are theories with only a free scalar in the bulk, one example of which we looked at in section \ref{sec:free}.  The subset of such free scalar theories appears to be very constrained \cite{Lauria:2020emq,Behan:2020nsf}, with essentially only the codimension $q=1$ case leading to defect theories which are not ``trivial''.  
\\

\noindent
Equally if not more interesting are defect theories with a free Maxwell field in the bulk.   Considerable research has been conducted on a codimension $q=1$ theory with a free photon in the bulk and charged fermionic matter on the boundary.  This theory is sometimes called mixed dimensional or reduced QED and has been used as an ``ultra-relativistic limit'' of graphene (see for example \cite{Kotikov:2013eha}).   
In section \ref{sec:free}, we looked at a $q=2$ and $q=3$ ``wedge'' theory as a prelude to looking at higher codimension theories
with charged matter on the defect.  
Literature suggests that the $q=2$ theory with charged matter on the defect
is problematic \cite{Gorbar:2001qt,Heydeman:2020ijz} because the effective photon propagator experienced by the matter has a logarithm in it and requires a scale to be well defined.  We would like to explore what happens in dimensional regularization, moving slightly away from the $q=2$ limit whether conformal defect constraints can be applied.  The $q=3$ case is also very interesting, for example in the study of Wilson and 't Hooft lines.

\section*{Acknowledgments}
We thank K.~Ray and V.~Schaub for discussion.  We also thank E.~Lauria, M.~Meineri, and E.~Trevisani for correspondence.  
This research was supported in part by the U.K.\ Science \& Technology Facilities Council Grants ST/P000258/1
and ST/T000759/1.
C.H.\ would like to acknowledge a Wolfson Fellowship from the Royal Society.

\appendix

\section{Useful Identities for $\langle J \mathcal{O} \rangle$, $\langle JJ \rangle$, $\langle T \mathcal{O} \rangle$,
and $\langle F \mathcal{O} \rangle$}
Here we list some identities used in deriving the constraints arising from the conservation equations. 

\subsection*{$\langle J \mathcal{O} \rangle$}

\begin{subequations}\label{JOidentities1}
	\begin{align}
	\partial_{\mu} \left(\frac{\Xi^{(1) \mu}}{(s^2)^{d-1}}\right) &=-\frac{2y'}{s^{2d}}(2\xi_1(q-1)+\xi_2d),\\
	\partial_{\mu} \left(\frac{\Xi^{(2) \mu}}{(s^2)^{d-1}}\right) &=\frac{2y'}{s^{2d}}(\xi_3(d-1)-2\xi_1(\xi_3/\xi_2+q-1)).
	\end{align}
\end{subequations}

\subsection*{$\langle JJ \rangle$}
\begin{equation} \label{JJidentities}
\begin{aligned}
\partial_{\mu}\Xi'^{(2)}_{\nu} &=\frac{2y'}{s^2} \left(2\xi_1/\xi_2 J'_{\mu \nu} \right), \\
\partial_{\mu} \Xi'^{(1)}_{\nu} &=-\frac{2y'}{s^2} I_{\mu \nu},\\
\end{aligned}
\qquad
\begin{aligned}
\partial_{\mu} \left(\frac{J'^{\mu}_{\nu}}{(s^2)^{d-1}}\right) &=\frac{2y'}{s^{2d}} \left(2\xi_1/\xi_2-(d-1)\right) \Xi'^{(2)}_{\nu}, \\
\partial_{\mu} \left(\frac{I^{\mu}_{\nu}}{(s^2)^{d-1}} \right) &=0,
\end{aligned}
\end{equation}
\subsection*{$\langle F \mathcal{O} \rangle$}
\begin{equation} \label{FOidentity}
\begin{aligned}
\partial_{\mu} \left(\frac{2}{s^4}\Xi^{(1)}_{[\mu} \Xi^{(2)}_{\nu]}   \right) = \frac{2|y'|}{s^6} &\bigg[\left(-3\xi_3 +2\xi_1 \left(\frac{\xi_3}{\xi_2} +q -1 \right)\right)\Xi^{(1)}_{\nu} \\
&+ \left(2\xi_1(2-q) +\xi_3 -2\xi_2\right)\Xi^{(2)}_{\nu} \bigg],
\end{aligned}
\end{equation}
\subsection*{$\langle T \mathcal{O} \rangle$}
\begin{equation} \label{TOidentities}
\begin{aligned}
\partial_{\mu} \left(\frac{1}{s^{2d}} \left(\Xi^{(1)\mu}\Xi^{(1)}_{\nu} -\frac{\delta^{\mu}_{\nu}}{u^2 d} \right) \right) &= \frac{2|y'|}{s^{2(d+1)}} \bigg[\left(2\xi_1(1-q)-\frac{\xi_2}{d}(d+2)(d-1)\right)\Xi^{(1)}_{\nu} \\
&+ \left(\frac{\xi_2}{d}(d-2)\right)\Xi^{(2)}_{\nu} \bigg], \\
\partial_{\mu} \left(\frac{1}{s^{2d}} \left(\Xi^{(2)\mu}\Xi^{(2)}_{\nu} -\frac{\delta^{\mu}_{\nu} \xi_3}{\xi_2 d} \right) \right) &= \frac{2|y'|}{s^{2(d+1)}} \bigg[\left(\frac{2\xi_1 \xi_3}{\xi_2}\right)\Xi^{(1)}_{\nu} \\
&+ \left(\xi_3d -4\xi_1\left(\frac{\xi_3}{\xi_2} -\frac{1}{\xi_2^2d}+\frac{q}{2}\right)  \right)\Xi^{(2)}_{\nu} \bigg], \\
\partial^{\mu} \left(\frac{1}{s^{2d}} \left(\Xi^{(1)}_{(\mu}\Xi^{(2)}_{\nu)} -\frac{\delta^{\mu}_{\nu} \xi_3}{2\xi_1 d} \right) \right) &= \frac{2|y'|}{s^{2(d+1)}} \bigg[ \\
&+\left( \frac{\xi_3}{2d}(d+1)(d-2) -\xi_1 \left(\frac{\xi_3}{\xi_2} +q -1 \right) \right)\Xi^{(1)}_{\nu}  \\
&+ \left(\frac{\xi_3}{2}-\xi_1 q -\frac{\xi_2}{d} \left(1 +\frac{d^2}{2}+\frac{1}{\xi_2^2}   \right)  \right)\Xi^{(2)}_{\nu} \bigg] \\
\partial^{\mu} \left(\frac{1}{s^{2d}} \left(J^{\mu}_{\nu} -\frac{(q-1)}{d}\delta^{\mu}_{\nu}  \right) \right) &=  \frac{2|y'|}{s^{2(d+1)}} \bigg[2\xi_1(q-1)\Xi^{(1)}_{\nu}+\xi_2d \Xi^{(2)}_{\nu}  \bigg],
\end{aligned}
\end{equation}
\section{Tensor Structures for $\langle TT \rangle$}
The correlation function between two different symmetric rank-2 tensors is given by,
\begin{equation} \label{symmetric2tensor}
\begin{aligned}
&\langle S_{\mu \nu}(x) S'_{\alpha\beta}(x') \rangle = \frac{|y'|^{\Delta-\Delta'}}{s^{2\Delta}} \bigg(f_1 I_{(\mu| (\alpha} I_{\beta) |\nu)} + h \mathcal{J}_{\mu \nu} \delta_{\alpha \beta} + h'' \mathcal{J}''_{\alpha \beta} \delta_{\mu \nu} + H\delta_{\mu \nu}\delta_{\alpha \beta}\\
&+ f_2 [I_{\mu \alpha}\mathcal{J}'_{\nu \beta} + I_{\mu \beta}\mathcal{J}'_{\nu \alpha} + I_{\nu \alpha}\mathcal{J}'_{\mu \beta}+ I_{\nu \beta}\mathcal{J}'_{\mu \alpha}]  + \sum_{n\geq m; s \geq r} f_{nmrs} \Xi^{(m)}_{(\mu} \Xi^{(n)}_{\nu)} \Xi'^{(r)}_{(\alpha} \Xi'^{(s)}_{\beta)}   \\
&+ f_{mr} \Xi^{(m)}_{(\mu} I_{\nu) (\alpha} \Xi'^{(r)}_{\beta)}+F_{mr}\Xi^{(m)}_{(\mu} \mathcal{J}'_{\nu) (\alpha}\Xi'^{(r)}_{\beta)} + G \mathcal{J}_{\mu \nu} \mathcal{J}''_{\alpha \beta} + \sum_{s\geq r}h'_{rs} \Xi'^{(r)}_{(\alpha} \Xi'^{(s)}_{\beta)} \delta_{\mu \nu}   \\
&+ \sum_{n\geq m}g_{mn}\Xi^{(m)}_{(\mu} \Xi^{(n)}_{\nu)} \mathcal{J}''_{\alpha \beta}+ \sum_{s\geq r}g'_{rs}\Xi'^{(r)}_{(\alpha} \Xi'^{(s)}_{\beta)} \mathcal{J}_{\mu \nu} +\sum_{n\geq m}h_{mn} \Xi^{(m)}_{(\mu} \Xi^{(n)}_{\nu)} \delta_{\alpha \beta} \\
&+ f \mathcal{J}'_{(\mu (\alpha|} \mathcal{J}'_{\nu) |\beta)} \bigg),
\end{aligned}
\end{equation}
where we use the summation convention except when explicitly restricting the sums so that $f_{mnrs}$ has 9 components, $g_{mn}$, $g'_{rs}$, $h_{mn}$ and $h'_{rs}$ each have 3 components. This is a total of 36 structures. 
\\

\noindent
The tensor structures appearing on the correlation function between two stress tensors \eqref{TTcorrelator} are the following,
\begin{equation} \label{q2structuresTT1}
	\begin{aligned}
	T^{(1)}_{\mu \nu;\alpha \beta} &= 2I_{\mu (\alpha}I_{\beta) \nu}, \\
	T^{(2)}_{\mu \nu;\alpha \beta} &= 4\Xi^{(1)}_{(\mu}I_{\nu) (\alpha}\Xi'^{(1)}_{\beta)}, \\
	T^{(3)}_{\mu \nu;\alpha \beta} &= 4\Xi^{(2)}_{(\mu}I_{\nu) (\alpha}\Xi'^{(2)}_{\beta)}, \\
	T^{(5)}_{\mu \nu;\alpha \beta} &=\Xi^{(1)}_{\mu}\Xi^{(1)}_{\nu}\Xi'^{(1)}_{\alpha}\Xi'^{(1)}_{\beta},\\
	T^{(6)}_{\mu \nu;\alpha \beta} &=\Xi^{(2)}_{\mu}\Xi^{(2)}_{\nu}\Xi'^{(2)}_{\alpha}\Xi'^{(2)}_{\beta},  
	\end{aligned}
	\qquad
	\begin{aligned}
	T^{(4)}_{\mu \nu;\alpha \beta} &=4\Xi^{(1)}_{(\mu}I_{\nu) (\alpha}\Xi'^{(2)}_{\beta)} + 4\Xi^{(2)}_{(\mu}I_{\nu) (\alpha}\Xi'^{(1)}_{\beta)} \\
	T^{(7)}_{\mu \nu;\alpha \beta} &=\Xi^{(1)}_{\mu}\Xi^{(1)}_{\nu}\Xi'^{(2)}_{\alpha}\Xi'^{(2)}_{\beta}  + \Xi^{(2)}_{\mu}\Xi^{(2)}_{\nu}\Xi'^{(1)}_{\alpha}\Xi'^{(1)}_{\beta}, \\
	T^{(8)}_{\mu \nu;\alpha \beta} &= 2\Xi^{(1)}_{\mu}\Xi^{(1)}_{\nu}\Xi'^{(1)}_{(\alpha}\Xi'^{(2)}_{\beta)} + 2 \Xi^{(1)}_{(\mu}\Xi^{(2)}_{\nu)}\Xi'^{(1)}_{\alpha}\Xi'^{(1)}_{\beta}, \\
	T^{(9)}_{\mu \nu;\alpha \beta} &= 2\Xi^{(2)}_{\mu}\Xi^{(2)}_{\nu} \Xi'^{(1)}_{(\alpha}\Xi'^{(2)}_{\beta)}  +2\Xi^{(1)}_{(\mu}\Xi^{(2)}_{\nu)} \Xi'^{(2)}_{\alpha}\Xi'^{(2)}_{\beta},\\
	T^{(10)}_{\mu \nu;\alpha \beta} &=4\Xi^{(1)}_{(\mu}\Xi^{(2)}_{\nu)}\Xi'^{(1)}_{(\alpha}\Xi'^{(2)}_{\beta)}.
	\end{aligned}
\end{equation}
\begin{equation}\label{q2structuresTT2}
\begin{aligned}
	S^{(1)}_{\mu \nu; \alpha \beta} &= 2\mathcal{J}'_{\mu (\alpha|}\mathcal{J}'_{\nu |\beta)} , \\
	S^{(2)}_{\mu \nu; \alpha \beta} &= 4\Xi^{(1)}_{(\mu}\mathcal{J}'_{\nu) (\alpha}\Xi'^{(1)}_{\beta)}, \\
	S^{(3)}_{\mu \nu; \alpha \beta} &= 4\Xi^{(2)}_{(\mu}\mathcal{J}'_{\nu) (\alpha}\Xi'^{(2)}_{\beta)}, \\
	S^{(9)}_{\mu \nu; \alpha \beta} &= 4I_{(\mu| (\alpha|}\mathcal{J}'_{|\nu) |\beta)},
	\end{aligned}
	\qquad
	\begin{aligned}
	S^{(4)}_{\mu \nu; \alpha \beta} &= 4\Xi^{(1)}_{(\mu}\mathcal{J}'_{\nu) (\alpha}\Xi'^{(2)}_{\beta)} + 4\Xi^{(2)}_{(\mu}\mathcal{J}'_{\nu) (\alpha}\Xi'^{(1)}_{\beta)}, \\
	S^{(5)}_{\mu \nu; \alpha \beta} &= \Xi^{(1)}_{\mu}\Xi^{(1)}_{\nu} \mathcal{J}''_{\alpha \beta} + \mathcal{J}_{\mu \nu} \Xi'^{(1)}_{\alpha} \Xi'^{(1)}_{\beta} , \\
	S^{(6)}_{\mu \nu; \alpha \beta} &= \Xi^{(2)}_{\mu}\Xi^{(2)}_{\nu} \mathcal{J}''_{\alpha \beta} + \mathcal{J}_{\mu \nu} \Xi'^{(2)}_{\alpha} \Xi'^{(2)}_{\beta}  , \\
	S^{(7)}_{\mu \nu; \alpha \beta} &= 2\Xi^{(1)}_{(\mu}\Xi^{(2)}_{\nu)} \mathcal{J}''_{\alpha \beta} + 2\mathcal{J}_{\mu \nu} \Xi'^{(1)}_{(\alpha} \Xi'^{(2)}_{\beta)} , \\
	S^{(8)}_{\mu \nu; \alpha \beta} &= \mathcal{J}_{\mu \nu} \mathcal{J}''_{\alpha \beta}.
	\end{aligned}
\end{equation}

\section{Taking the Defect Limit of $\langle V \mathcal{O}\rangle$, $\langle V V' \rangle$, $\langle S \mathcal{O} \rangle$, $ \langle F V' \rangle$ and $\langle FF' \rangle$} \label{bulktodefectlimitappendix}
Here we list some of the bulk-bulk correlation functions used for obtaining the bulk-defect correlations function. The only difference here compared to section \ref{bulktwopointcorrelation} is the choice of overall normalisation and the tensor structure basis as mentioned in section \ref{sec:bulktodefectlimit}. 

\subsection*{$\langle V \mathcal{O}\rangle$} 
The correlation function between any vector (no need to be conserved) and a scalar is given by,
\begin{equation}\label{VObulk}
\langle V_{\mu} (x) \mathcal{O}(x') \rangle = \frac{|y|^{\Delta' - \Delta}}{(s^2)^{\Delta'}} \left(g_1 \Xi^{(1)}_{\mu} + g_2 \frac{\xi_2}{\xi_1} \Xi^{(2)}_{\mu} \right).
\end{equation}

\subsection*{$\langle V V' \rangle$} 
The correlation function between any two distinct vectors is given by,
\begin{equation}\label{VVbulk}
\begin{split}
\langle V_{\mu}(x) V'_{\nu}(x') \rangle = \frac{|y|^{\Delta'-\Delta}}{(s^2)^{\Delta'}} &\bigg(g_1  \Xi^{(1)}_{\mu} \mathcal{X}'_{\nu} + g_2 \frac{\xi_2^2}{\xi_1^3}\Xi^{(2)}_{\mu} \Xi'^{(2)}_{\nu} + g_3 \frac{\xi_2}{\xi_1^2}\Xi^{(1)}_{\mu} \Xi'^{(2)}_{\nu} \\
\qquad
&+ g_4 \frac{\xi_2}{\xi_1}\Xi^{(2)}_{\mu} \mathcal{X}'_{\nu}+g_5\mathcal{I}_{\mu \alpha} + g_6  \bar{\mathcal{J}}'_{\mu \nu} \bigg).
\end{split}
\end{equation}

\subsection*{$\langle S \mathcal{O} \rangle$} 
The correlation function between any symmetric, traceless tensor and a scalar is given by,
\begin{equation} \label{SObulk}
\begin{split}
\langle S_{\mu \nu}(x) \mathcal{O}(x') \rangle = \frac{|y|^{\Delta'-\Delta}}{(s^{2})^{\Delta'}} &\bigg[g_1\left(\Xi^{(1)}_{\mu}\Xi^{(1)}_{\nu}-\frac{\delta_{\mu \nu}}{u^2 d}\right) + g_2\left(\frac{\xi_2^2}{\xi_1^2}\Xi^{(2)}_{\mu}\Xi^{(2)}_{\nu} - \frac{\delta_{\mu \nu} \xi_2 \xi_3}{\xi_1^2 d}\right) \\
\qquad
+&g_3\left(\frac{\xi_2}{\xi_1}\Xi^{(1)}_{(\mu}\Xi^{(2)}_{\nu)} + \frac{\delta_{\mu \nu} \xi_2\xi_3}{2\xi_1^2 d} \right) + g_4\left(\mathcal{J}_{\mu\nu}-\frac{q-1}{d}\delta_{\mu \nu} \right) \bigg].
\end{split}
\end{equation}

\subsection*{$ \langle F V' \rangle$}
The correlation function between any antisymmetric tensor and a vector is given by,
\begin{equation} \label{FVbulk}
\begin{split}
\langle F_{\mu \nu}(x) V'_{\alpha}(x') \rangle =  \frac{|y|^{\Delta'-\Delta}}{(s^{2})^{\Delta'}} &\bigg(g_1 \frac{\xi_2}{\xi_1}\Xi^{(1)}_{[\mu} \Xi^{(2)}_{\nu]} \mathcal{X}'_{\alpha} + g_2 \frac{\xi_2^2}{\xi_1^3} \Xi^{(1)}_{[\mu} \Xi^{(2)}_{\nu]} \Xi'^{(2)}_{\alpha} + 2g_3 \Xi^{(1)}_{[\mu} \mathcal{I}^{\phantom{(1)}}_{\nu] \alpha}\\
\qquad
&+ g_4 \frac{\xi_2}{\xi_1}\Xi^{(2)}_{[\mu} \mathcal{I}^{\phantom{(1)}}_{\nu] \alpha} + g_5 \Xi^{(1)}_{[\mu} \bar{\mathcal{J}}'_{\nu] \alpha} + g_6 \frac{\xi_2}{\xi_1}\Xi^{(2)}_{[\mu} \bar{\mathcal{J}}'_{\nu] \alpha} \bigg).
\end{split} 
\end{equation}

\subsection*{$\langle F F' \rangle$}
The correlation function between any two different antisymmetric tensors is given by,
\begin{equation}  \label{FFbulk}
\begin{split}
\langle F_{\mu \nu}(x) &F'_{\alpha \beta}(x') \rangle = \frac{4|y|^{\Delta'-\Delta}}{(s^2)^{\Delta'}} \bigg[g_1 \mathcal{I}_{\mu [\alpha|} \mathcal{I}_{\nu |\beta]} + g_2 \Xi^{(1)}_{[\nu} \mathcal{I}^{\phantom{(1)}}_{\mu] [\alpha} \mathcal{X}'_{\beta]} \\
+ &g_3\frac{\xi_2}{\xi_1^2}\Xi^{(1)}_{[\nu} \mathcal{I}^{\phantom{(1)}}_{\mu] [\alpha} \Xi'^{(2)}_{\beta]} + g_4 \frac{\xi_2}{\xi_1} \Xi^{(2)}_{[\nu} \mathcal{I}^{\phantom{(1)}}_{\mu] [\alpha} \mathcal{X}'_{\beta]}  + g_5 \frac{\xi_2^2}{\xi_1^3}\Xi^{(2)}_{[\nu} \mathcal{I}^{\phantom{(1)}}_{\mu] [\alpha} \Xi'^{(2)}_{\beta]}  \\
+ &g_6 \frac{\xi_2^2}{\xi_1^3} \Xi^{(1)}_{[\mu}\Xi^{(2)}_{\nu]} \mathcal{X}'_{[\alpha} \Xi'^{(2)}_{\beta]} + \frac{g_7}{2} \bar{\mathcal{J}}'_{\mu [\alpha|} \bar{\mathcal{J}}'_{\nu |\beta]} + g_8 \Xi^{(1)}_{[\mu} \bar{\mathcal{J}}'_{\nu] [\beta}\mathcal{X}'_{\alpha]} + g_9 \frac{\xi_2}{\xi_1^2} \Xi^{(1)}_{[\mu} \bar{\mathcal{J}}'_{\nu] [\beta} \Xi'^{(2)}_{\alpha]} \\
+ &g_{10} \frac{\xi_2}{\xi_1} \Xi^{(2)}_{[\mu} \bar{\mathcal{J}}'_{\nu] [\beta} \mathcal{X}'_{\alpha]} + g_{11} \frac{\xi_2^2}{\xi_1^3} \Xi^{(2)}_{[\mu}\bar{\mathcal{J}}'_{\nu] [\beta}\Xi'^{(2)}_{\alpha]} + g_{12}\bar{\mathcal{J}}'_{[\mu [\alpha|} \mathcal{I}_{\nu] |\beta]} \bigg].
\end{split}
\end{equation}
\section{$\langle TT \rangle $ in the Free Maxwell Theory for $d=4$, $q=2$}
Given that the two point correlation function of the field strength $F_{\mu\nu}$ obeys the defect conformal symmetry \eqref{generalFF},
\begin{equation} \label{appcorrelFF}
\begin{split}
\langle F_{\mu \nu}(x) &F_{\alpha \beta}(x') \rangle = \frac{4}{(s^2)^2} \bigg[f_1 I_{[\mu| [\alpha}I_{\beta] |\nu]} + f_2 \Xi^{(1)}_{[\nu} I_{\mu] [\alpha} \Xi'^{(1)}_{\beta]} \\
+& f_3\left( \Xi^{(1)}_{[\nu} I^{\phantom{(1)}}_{\mu] [\alpha} \Xi'^{(2)}_{\beta]} + \Xi^{(2)}_{[\nu} I^{\phantom{(1)}}_{\mu] [\alpha} \Xi'^{(1)}_{\beta]} \right) + f_4\Xi^{(2)}_{[\nu} I^{\phantom{(1)}}_{\mu] [\alpha} \Xi'^{(2)}_{\beta]} + f_5 \Xi^{(1)}_{[\nu}\Xi^{(2)}_{\mu]} \Xi'^{(1)}_{[\alpha} \Xi'^{(2)}_{\beta]} \bigg],
\end{split}
\end{equation}
and given the bulk Maxwell energy momentum tensor $T_{\mu \nu}(x)=\tensor{F}{_\mu ^\gamma}\tensor{F}{_\nu _\gamma} -\frac{\delta_{\mu\nu}}{4} F_{\alpha \beta}F^{\alpha \beta}$, we find the two point correlation function of $T_{\mu\nu}$ is given by \eqref{TTcorrelator},
\begin{equation} \label{TTcorrelatorappendix}
\begin{split}
\langle T_{\mu \nu}(x) &T_{\alpha \beta}(x') \rangle = \frac{1}{s^{8}}\bigg[\sum_{n=1}^{10}g_n(\xi_1,\xi_2)T^{(n)}_{\mu \nu; \alpha \beta} +h_1 \delta_{\mu \nu} \delta_{\alpha \beta}  \\
&+ h_2 \left(\delta_{\mu \nu}\Xi'^{(1)}_{\alpha}\Xi'^{(1)}_{\beta} + \delta_{\alpha \beta}\Xi^{(1)}_{\mu}\Xi^{(1)}_{\nu} \right) + h_3 \left(\delta_{\mu \nu}\Xi'^{(2)}_{\alpha}\Xi'^{(2)}_{\beta} + \delta_{\alpha \beta}\Xi^{(2)}_{\mu}\Xi^{(2)}_{\nu} \right)  \\
&+h_4 \left( \delta_{\mu \nu}(\Xi'^{(1)}_{\alpha}\Xi'^{(2)}_{\beta} + \Xi'^{(1)}_{\beta}\Xi'^{(2)}_{\alpha}) + \delta_{\alpha \beta}(\Xi^{(1)}_{\mu}\Xi^{(2)}_{\nu} + \Xi^{(1)}_{\nu}\Xi^{(2)}_{\mu})\right)  \bigg],
\end{split}
\end{equation}
where, 
\begin{equation}
	\begin{split}
	g_1 &= (d-2)f_1^2 +\frac{f_2^2}{u^4} + \frac{\xi_3^2}{\xi_2^2}f_4^2 +\xi_3\left(\frac{2}{\xi_2 u^2} +\frac{\xi_3}{2\xi_1^2}\right)f_3^2 +2\xi_2 \left(\frac{1}{u^2} +\frac{\xi_3}{2\xi_1}\right)f_1 f_2  \\
	&+2\xi_3 \left(\frac{\xi_3}{2\xi_1} -1\right)f_1 f_4 +\frac{\xi_3^2}{2\xi_1^2}f_2 f_4 -4\xi_3 \left(\frac{\xi_2}{2\xi_1} +1 \right)f_1 f_3 -\frac{2\xi_3}{\xi_1 u^2}f_2 f_3 -\frac{2\xi_3^2}{\xi_1 \xi_2}f_3 f_4,  \\
	g_2 &= \xi_2\left(\frac{\xi_3}{2\xi_1} - \frac{1}{u^2}\right)f_2^2 -\xi_3 \left(2+\frac{\xi_3}{\xi_1}\right)f_3^2 +(d-3-\xi_2^2)f_1 f_2 -\xi_3^2 f_1 f_4 \\
	&+\xi_3 \left(\frac{\xi_3}{2\xi_1} -1\right)f_1 f_5 -\xi_3 \left(1+\frac{\xi_3}{2\xi_1}\right)f_2 f_4 +\frac{\xi_3^2}{4\xi_1^2}f_2 f_5 +\frac{\xi_3^2}{\xi_2^2}f_4 f_5 +2\xi_2 \xi_3 f_1 f_3  \\
	&+\frac{3\xi_2\xi_3}{\xi_1}f_2f_3 +\frac{2\xi_3^2}{\xi_2}f_3f_4 -\frac{\xi_3^2}{\xi_1\xi_2}f_3f_5 , \\
	g_3 &= \xi_3 \left(\frac{\xi_3}{2\xi_1}+1\right)f_4^2 + \xi_2 \left(\frac{2}{u^2} -\frac{\xi_3}{\xi_1}\right)f_3^2 -\xi_2^2 f_1 f_2 +(d-3-\xi_2^2)f_1 f_4 \\
	&+\xi_2 \left(\frac{1}{u^2} +\frac{\xi_3}{2\xi_1}\right)f_1 f_5 +\xi_2 \left(\frac{1}{u^2} -\frac{\xi_3}{2\xi_1}\right)f_2 f_4 +\frac{f_2 f_5}{u^4} + \frac{\xi_3^2}{4 \xi_1^2}f_4 f_5 -2\xi_2^2 f_1 f_3  \\
	&+ \frac{2\xi_2}{u^2}f_2 f_3 - \frac{3\xi_2 \xi_3}{\xi_1}f_3 f_4 -\frac{\xi_3}{\xi_1 u^2}f_3 f_5, \\
	g_4 &= \frac{\xi_2}{u^2}f_2^2 + \frac{\xi_3^2}{\xi_2}f_4^2 +\xi_2^2 f_1 f_2 - \xi_2\xi_3 f_1 f_4 +\xi_3 \left(\frac{\xi_2}{2\xi_1}+1\right)f_1 f_5 +\frac{\xi_3}{2\xi_1 u^2}f_2 f_5 \\
	&+ \frac{\xi_3^2}{2\xi_1\xi_2}f_4 f_5 +(d-3+\xi_2^2-\xi_2\xi_3)f_1f_3 + \xi_2\left(\frac{1}{u^2} -\frac{\xi_3}{2\xi_1}\right)f_2f_3  \\
	&-\xi_3 \left(1+\frac{\xi_3}{2\xi_1}\right)f_3f_4 -\xi_3 \left(\frac{\xi_3}{4\xi_1^2} + \frac{1}{\xi_2 u^2}\right)f_3f_5,  \\
	g_5 &= 2(d-4+2\xi_2^2)f_2^2 +\frac{2\xi_3^2}{\xi_2^2}f_5^2 + 4\xi_3^2 f_3^2 -4\xi_3^2 f_1 f_5 -2\xi_3 \left(\frac{\xi_3}{\xi_1} +2\right)f_2f_5 \\
	&-8\xi_2 \xi_3 f_2 f_3  + \frac{8 \xi_3^2}{\xi_2}f_3 f_5,  \\
	g_6 &= \frac{2f_5^2}{u^4} +2(d-4+2\xi_2^2)f_4^2 +4\xi_2^2 f_3^2 -4\xi_2^2f_1 f_5 +2\xi_2\left(\frac{2}{u^2} -\frac{\xi_3}{\xi_1}\right)f_4f_5 \\
	&+8\xi_2^2 f_3 f_4 + \frac{8\xi_2}{u^2}f_3 f_5 , \\
	g_7 &= \frac{2\xi_3}{\xi_2 u^2} f_5^2 + 2(d-4+2\xi_2 \xi_3)f_3^2 -4\xi_2^2 f_1 f_5 -4\xi_2^2 f_2 f_4 -\frac{4\xi_2}{u^2}f_2 f_5 +4\xi_3 f_4 f_5 \\
	&-4\xi_2^2f_2f_3 +4\xi_2\xi_3 f_3f_4 +4\xi_3 \left(1+\frac{\xi_2}{2\xi_1}\right)f_3f_5,  \\
		g_8 &= -2\xi_2^2 f_2^2 +\frac{\xi_3^2}{\xi_1\xi_2}f_5^2 -4\xi_2\xi_3 f_1f_5 -2\xi_2\xi_3 f_2 f_4 - \frac{3\xi_2 \xi_3}{\xi_1} f_2 f_5 +\frac{2\xi_3^2}{\xi_2}f_4f_5 +2\xi_3^2 f_3 f_4\\
	&+2(d-3-\xi_2^2)f_2f_3 +2\xi_3 \left(1+\frac{\xi_3}{2\xi_1}\right)f_3 f_5, \\
		g_9 &= \frac{\xi_3}{\xi_1 u^2} f_5^2 + 2\xi_2\xi_3 f_4^2 +4\xi_2^2 f_1 f_5 +2\xi_2^2 f_2 f_4 +\frac{2\xi_2}{u^2}f_2 f_5 +\frac{3\xi_2 \xi_3}{\xi_1}f_4 f_5 +2\xi_2^2f_2 f_3 \\
	&+2(d-3-\xi_2^2)f_3f_4 +2\xi_2 \left(\frac{\xi_3}{2\xi_1} -\frac{1}{u^2}\right)f_3 f_5, 
	\end{split}
	\end{equation}
	\begin{equation*}
\begin{split}
	g_{10} &= \xi_2^2 f_2^2 +\frac{\xi_3^2}{2\xi_1^2}f_5^2 +\xi_3^2 f_4^2 +(4-d)f_3^2 -2\xi_2(\xi_2-\xi_3)f_1f_5 + (d-4+2\xi_2\xi_3)f_2f_4  \\ 
	&+ \xi_2 \left( \frac{3\xi_3}{2\xi_1}-\frac{1}{u^2} \right) f_2f_5 + \xi_3 \left(\frac{3\xi_3}{2\xi_1} +1\right)f_4 f_5 + \xi_3 \left(\frac{\xi_2}{\xi_1} -\frac{2}{u^2}\right)f_3 f_5,
	\end{split}
\end{equation*}
and the functions $h_1,...,h_4$ are obtained from the traceless condition \eqref{TTconstraints}.

\bibliographystyle{JHEP}
\bibliography{Defect}

\end{document}